\documentclass[preprint]{aastex631}
\usepackage{epsfig,psfig}

\newcolumntype{C}[1]{>{\centering\arraybackslash}p{#1}}
\newcommand{\E}[1]{\times 10^{#1}}
\newcommand{\s}{\,{\rm s}}      \newcommand{\ps}{\,{\rm s}$^{-1}$}
\newcommand{\yr}{\,{\rm yr}}    \newcommand{\Msun}{\,{\rm M}_{\rm \odot}}
\newcommand{\cm}{\,{\rm cm}}    \newcommand{\km}{\,{\rm km}}
\newcommand{\parsec}{\,{\rm pc}}\newcommand{\kpc}{\,{\rm kpc}}
        \newcommand{\K}{\,{\rm K}}
\newcommand{\amin}{\,{\rm arcmin}} 

\newcommand{\ncol}{$N{\rm (H_2})$}

\newcommand{\degree}{$^\circ$}

\newcommand{\HII}{H~{\sc ii}}
       
\newcommand{\twCO}{$^{12}$CO}  \newcommand{\thCO}{$^{13}$CO}
\newcommand{\CeiO}{C$^{18}$O}
\newcommand{\myemail}{xinzhou@pmo.ac.cn}
\shorttitle{Properties of MCs in the Milky Way}
\shortauthors{Zhou et al.}

\begin{document}

\title{Statistical Properties of Molecular Clouds in the Milky Way: Insights from Three-Isotopologue CO Observations of the MWISP Project
}

\correspondingauthor{Xin Zhou}
\email{\myemail}
\author{Xin Zhou}
\affiliation{Purple Mountain Observatory, Chinese Academy of Sciences, 10 Yuanhua Road, Nanjing 210033, People's Republic of China}
\affiliation{State Key Laboratory of Radio Astronomy and Technology, Purple Mountain Observatory, Chinese Academy of Sciences, 10 Yuanhua Road, Nanjing 210023, China}
\author{Ji Yang} 
\affiliation{Purple Mountain Observatory, Chinese Academy of Sciences, 10 Yuanhua Road, Nanjing 210033, People's Republic of China}
\affiliation{State Key Laboratory of Radio Astronomy and Technology, Purple Mountain Observatory, Chinese Academy of Sciences, 10 Yuanhua Road, Nanjing 210023, China}
\author{Qing-Zeng Yan}
\affiliation{Purple Mountain Observatory, Chinese Academy of Sciences, 10 Yuanhua Road, Nanjing 210033, People's Republic of China}
\author{Yan Sun}
\affiliation{Purple Mountain Observatory, Chinese Academy of Sciences, 10 Yuanhua Road, Nanjing 210033, People's Republic of China}
\affiliation{State Key Laboratory of Radio Astronomy and Technology, Purple Mountain Observatory, Chinese Academy of Sciences, 10 Yuanhua Road, Nanjing 210023, China}
\author{Lixia Yuan}
\affiliation{Purple Mountain Observatory, Chinese Academy of Sciences, 10 Yuanhua Road, Nanjing 210033, People's Republic of China}
\author{Yang Su}
\affiliation{Purple Mountain Observatory, Chinese Academy of Sciences, 10 Yuanhua Road, Nanjing 210033, People's Republic of China}
\affiliation{State Key Laboratory of Radio Astronomy and Technology, Purple Mountain Observatory, Chinese Academy of Sciences, 10 Yuanhua Road, Nanjing 210023, China}
\author{Xuepeng Chen}
\affiliation{Purple Mountain Observatory, Chinese Academy of Sciences, 10 Yuanhua Road, Nanjing 210033, People's Republic of China}
\affiliation{State Key Laboratory of Radio Astronomy and Technology, Purple Mountain Observatory, Chinese Academy of Sciences, 10 Yuanhua Road, Nanjing 210023, China}
\author{Shaobo Zhang}
\affiliation{Purple Mountain Observatory, Chinese Academy of Sciences, 10 Yuanhua Road, Nanjing 210033, People's Republic of China}
\affiliation{State Key Laboratory of Radio Astronomy and Technology, Purple Mountain Observatory, Chinese Academy of Sciences, 10 Yuanhua Road, Nanjing 210023, China}

\begin{abstract}
We present a comprehensive statistical analysis of molecular cloud (MC) properties using the MWISP survey's \twCO, \thCO, and \CeiO~(J=1--0) data toward the inner ($l$=45\degree--60\degree) and outer ($l$=120\degree--130\degree) Galaxy. From a strict selection of 24,724 identified MCs, a final sample of 3,161 well-resolved MCs is established. 
We investigate the distributions of observational, morphological, and derived physical parameters, as well as their environmental dependencies and intercorrelations.
Our analysis reveals that MCs are typically oblate and tend to align with the Galactic disk.
A critical evaluation using a nearby subsample confirms significant distance-dependent selection effects for some parameters, nevertheless, the direction of changes in these parameters can indicate distance influence.
We also examine several specific subsamples, revealing the distinct characteristics of MCs in the G120 spiral shock region, MCs in the G50 interarm spurs, \CeiO-bright MCs, and MCs with supra-Larson velocity dispersion. For instance, MCs with supra-Larson velocity dispersion are predominantly small and likely young clouds inheriting turbulence from the diffuse ISM.
Notably, a comparison across tracers reveals that typical MCs have a turbulent, diffuse, \twCO-bright gas structure in their outer layers that does not contribute directly to star formation. In contrast, \thCO-bright gas represents a turning point where gravity becomes significant; \CeiO-bright gas is about gravity-dominated. 
Comprehensive correlation analysis confirms a flatter $\sigma_{\rm v}$-size relation than classic Larson's law and a strong mass–size relation. Incorporating dimensional analysis, we derive minimal sets of eigenparameters from which most other observational and physical parameters can be estimated. This highlights the underlying scaling relations that governing cloud properties.

\end{abstract}



\section{Introduction} \label{sec:intro}
Molecular clouds (MCs) are the primary sites of star and planet formation.
Their mass and distribution are closely related to the overall star formation activity in a galaxy \citep{SolomonVandenbout2005}.
A comprehensive understanding of MC properties, such as their density, temperature, and velocity structure, is therefore a fundamental step in investigating the mechanisms of star formation and galactic evolution.

The study of MCs relies heavily on molecular line tracers. Although molecular hydrogen is the dominant component, it is difficult to be observed directly due to its lack of a permanent dipole moment. Carbon monoxide (CO), particularly its isotopes, is widely employed as a tracer for molecular gas. 
The three CO isotopic lines, \twCO, \thCO, and \CeiO~(J=1--0), have different optical depths and critical densities, which allow us to probe various physical conditions and layers within MCs. 
Numerous large-scale Galactic plane surveys of these lines have been performed, e.g., the Columbia CO survey \citep{Dame+2001, DameThaddeus2022}, the NANTEN CO survey \citep{MizunoFukui2004}, the Galactic Ring Survey \citep[GRS;][]{Jackson+2006}, the FOREST Unbiased Galactic plane Imaging survey \citep[FUGIN;][]{Umemoto+2017}, and the Mopra CO survey \citep{Burton+2013, Braiding+2015, Braiding+2018, Cubuk+2023}, among others. These surveys have significantly advanced our knowledge of molecular gas distribution and cloud properties.
Nevertheless, some key questions still persist: How do the physical states (e.g., turbulence, gravitational binding, and pressure) of molecular gas traced by these different lines differ systematically? Are these differences important in the evolution of clouds or the formation of stars?

MCs exhibit remarkable diversity in size, shape, mass, and density, spanning several orders of magnitude \citep[e.g.,][]{Yuan+2021, Neralwar+2022a, Clarke+2022}. Their evolution connects galactic-scale dynamics with small-scale star formation processes. Key physical properties of molecular gas, such as surface mass density, exhibit enhancements along spiral arms \citep{Roman-Duval+2010} and demonstrate systematic trends with Galactic radius \citep{Sun+2024s, Zhou+2025}. 
The internal state of MCs is governed by a combination of thermal, turbulent, and magnetic pressures. 
Turbulence plays a vital role in the interstellar medium (ISM) across different scales, providing pressure support within MCs against gravitational collapse, regulating mass and energy distribution, and influencing the timescales for gas cooling and chemical evolution. 
This is reflected in scaling relations, such as the size–line width relation proposed by \cite{Larson1981}. 
Magnetic fields provide additional pressure support that influences gas flow, angular momentum transfer, and collapse geometry.
The contribution of magnetic fields varies at different scales \citep[see][for reference]{Crutcher2012}.

There are well-documented correlations among various MC parameters, such as size, mass, density, velocity dispersion, and magnetic field strength \citep[e.g.,][]{Larson1981, Solomon+1987, Heyer+2009, Ballesteros-Paredes+2011, Beaumont+2012, Shetty+2012, CaselliMyers1995, Sun+2020, Huang+2023, Cahlon+2024}.
These correlations are significant for understanding the evolution and star formation of MCs.
Some of these relationships can be explained by turbulent models \citep[e.g.,][]{FleckRobert1996, Vazquez-Semadeni+1997}, which have profound implications for cloud evolution and star formation efficiency. Density is a particularly critical parameter, as evidenced by its correlation with star formation rate \citep[e.g., shown by the Gao-Solomon relation, etc.][]{GaoSolomon2004, Lada+2010, Neumann+2024}.
However, significant questions remain about the formation mechanism of MCs and the role of galactic dynamics.
The statistical properties of MCs and the correlations among their parameters vary in different galactic environments, such as the inner versus outer galaxy or spiral arms versus interarm regions, which are not well studied.
More observational data and more detailed analysis are needed to further clarify these questions.

The primary objective of this study is to conduct a systematic, multi-tracer census of MC properties toward specific regions of the inner (G50) and outer (G120) Galaxy. Utilizing the unbiased \twCO, \thCO, and \CeiO~(J=1--0) data from the Milky Way Imaging Scroll Painting (MWISP) survey, we aim to characterize the statistical distributions of cloud properties, compare the physical conditions traced by the three CO isotopic lines, and analyze parameter correlations to investigate the underlying scaling relations.

We describe the data and MC identification in Section~\ref{sec:data}. Section~\ref{sec:result} presents the results, including a catalog of well-resolved MCs and their derived properties (Section~\ref{sec:para}), the statistical distributions of their parameters (Section~\ref{sec:paradist}), and the correlations between different parameters (Section~\ref{sec:paracorr}). 
Section~\ref{sec:paradist} is divided into several parts: an overview of parameter distributions (Section~\ref{sec:distover}), the effect of distance on parameter distributions (Section~\ref{sec:disteff}), the properties of specific subsamples (Section~\ref{sec:subsampprop}), and a comparison across molecular tracers (Section~\ref{sec:tracercomp}).
Our conclusions are summarized in Section~\ref{sec:sum}.

\section{Data and Molecular Cloud Sample} \label{sec:data}
The CO spectral line data used in this study are obtained from the Milky Way Imaging Scroll Painting (MWISP\footnote{http://www.radioast.nsdc.cn/mwisp.php}) project\citep{Yang+2026}, which was conducted using the 13.7-meter millimeter-wavelength telescope at the Purple Mountain Observatory (PMO) in Delingha, China. 
We focus on two regions toward the inner (G50: $l = 44.75^\circ$–$60.25^\circ$) and outer (G120: $l = 119.75^\circ$–$130.25^\circ$) Galaxy. Both of the two regions have the Galactic latitudes of $|b|\le5.25$\degree.
The \twCO, \thCO, and \CeiO\ (J=1--0) lines were mapped simultaneously.
The half-power beamwidth is $\sim$51$''$, with spectral resolutions of 0.17~\km\ps\ for \twCO\ and 0.16~\km\ps\ for \thCO/\CeiO.
The data were reduced and gridded with 30$''$ spacing using the standard MWISP pipeline and the GILDAS/CLASS package\footnote{http://www.iram.fr/IRAMFR/GILDAS}.
The final data cubes have typical RMS noise levels of $\sim$0.45~K per channel for \twCO\ and $\sim$0.24~K per channel for \thCO\ and \CeiO.
We performed a comprehensive analysis of the molecular gas distribution using the same dataset \citep{Zhou+2025}, ensuring consistency across our research framework.

\begin{figure*}[ptbh!]
\centerline{{\hfil\hfil
\psfig{figure=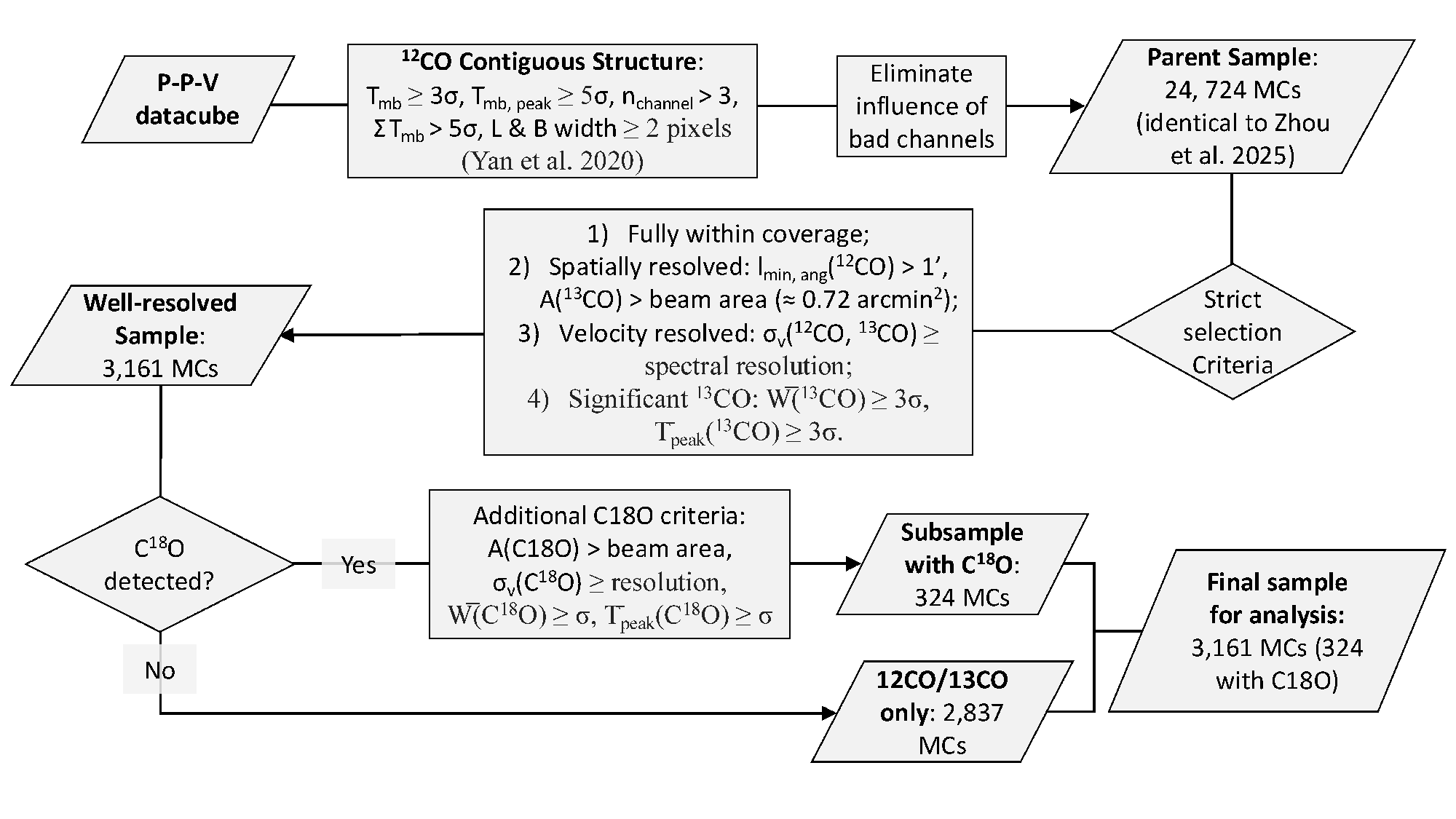,width=6in,angle=0, clip=}
\hfil\hfil}}
\caption{Flowchart illustrating the method for constructing MC samples.
}
\label{f:flowchart}
\end{figure*}

\begin{figure*}[ptbh!]
\centerline{{\hfil\hfil
\psfig{figure=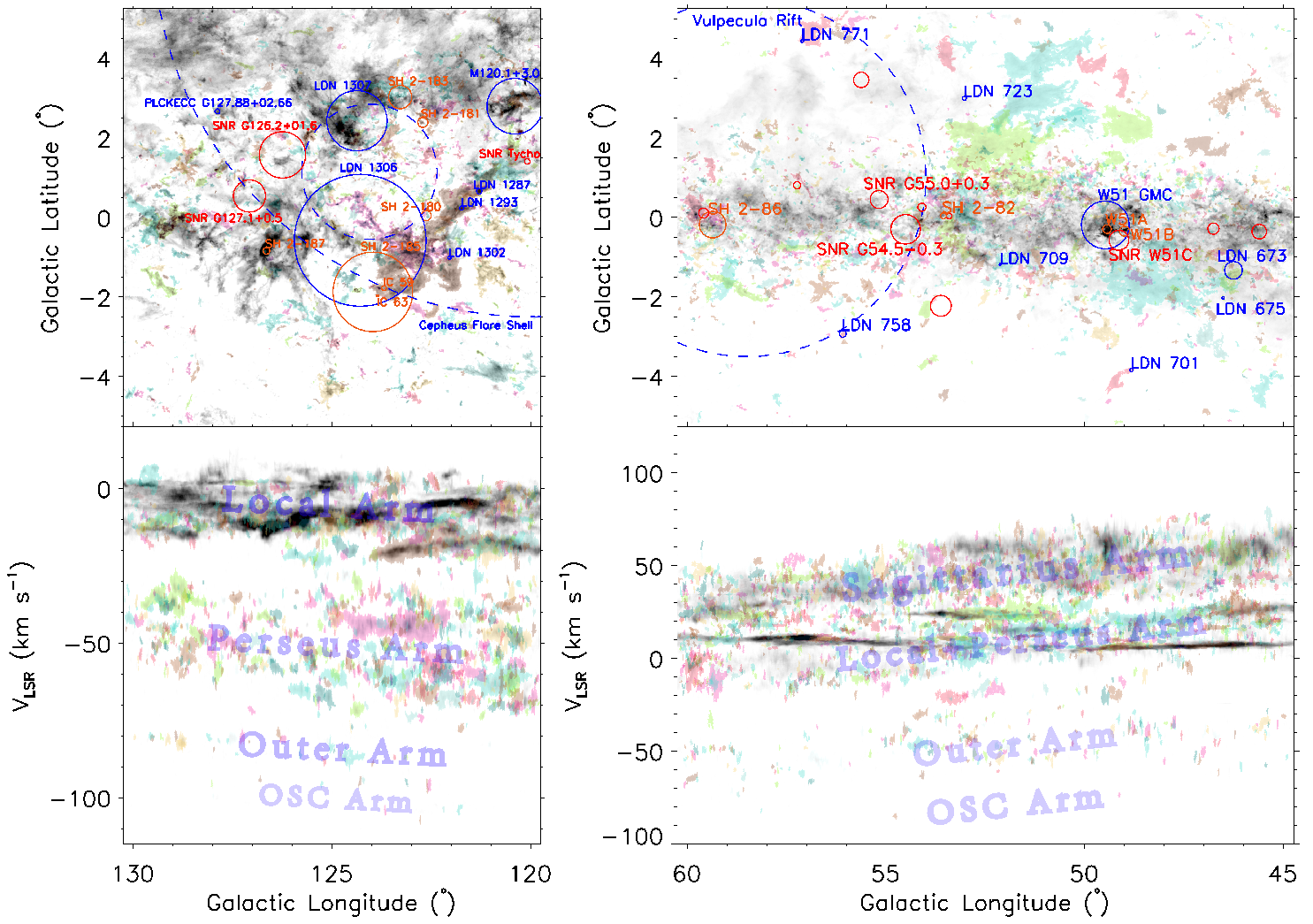,width=6in,angle=0, clip=}
\hfil\hfil}}
\caption{\twCO\ (J=1--0) integrated intensity map, along with corresponding Galactic longitude-velocity ($l$-$v$) map, for the G120 (left) and G50 (right) regions. 
The MCs studied here are overlaid in color, with each cloud shown by a distinct random color.
The data were initially moment-masked \citep[see][for reference]{Dame2011} to suppress noise while integrating over large velocity ranges. The minimum values of the maps are zero, and the intensity is on a linear scale. The maximum values of the integrated intensity maps are 41.9 and 166.1~K\,\km\ps\ for the G120 and G50 regions, respectively. Integrated intensities greater than these maxima are truncated for better visibility. 
Some well-studied objects in the region, including supernova remnants (red), \HII\ regions (orange), and others (blue), are indicated by circles \citep[see][and references therein, for details]{Zhou+2025}. 
The approximate locations of the different spiral arms are also indicated on the $l$-$v$ map.
}
\label{f:map}
\end{figure*}

Figure~\ref{f:flowchart} presents a flowchart illustrating the method for constructing MC samples.
A catalog of MCs is firstly constructed from the \twCO~(J=1--0) data.
Here, an MC is defined here as a spatially and spectrally contiguous structure within the position-position-velocity data cube. 
The minimum threshold for MCs is 3$\sigma$ in the main-beam temperature ($T_{\rm mb}$), with some faint signals excluded.
A selection process is applied, building upon the methodologies established in related works \citep[e.g.,][]{Yan+2020, Zhou+2025}, to select significant and resolved structures suitable for kinematic and physical analysis. 
The parent MC sample (24,724 MCs) is the same as that built by \citep{Zhou+2025}.
To ensure the physical reliability of the extracted clouds and the accuracy of their derived properties, we applied additional strict criteria: 
(1) fully contained within the observation coverage; 
(2) spatially resolved, with a \twCO\ angular length (full width at half maximum, FWHM) of the minor axis, $l_{\rm min,\,ang}$(\twCO) $> 1'$ and an angular area of \thCO, $A$(\thCO) greater than the squared beamwidth $\approx 0.72~\amin^2$;
(3) resolved in velocity, with \twCO\ and \thCO\ velocity dispersions, $\sigma_{\rm v}$(\twCO,\,\thCO) greater than the corresponding spectral resolutions;
(4) \thCO\ average integrated intensity $\overline{W}({\rm ^{13}CO}) \ge 3\sigma$ and \thCO\ peak temperature of average spectrum $\overline{T}_{\rm peak}(^{13}{\rm CO}) \ge 3\sigma$;
and (5) for parameters related to \CeiO\ emission, \CeiO\ angular area $A$(\CeiO) greater than the squared beamwidth $\approx 0.72~\amin^2$, \CeiO\ velocity dispersion $\sigma_{\rm v}$(\CeiO) exceeds the \CeiO\ velocity resolution, \CeiO\ average integrated intensity $\overline{W}$(\CeiO) $\ge \sigma$, and \CeiO\ peak temperature of average spectrum $\overline{T}_{\rm peak}$(\CeiO) $\ge \sigma$. 
In contrast to the voxel-level identification of the \CeiO\ core adopted by \cite{Yang+2026}, which yields a significant sample of \CeiO\ cores, our approach considers the integrated \CeiO\ emission across the entire cloud region, and therefore identifies a larger sample of \CeiO\ clouds.
The final catalog consists of 3,161~MCs, which we will refer to as the well-resolved sample. This sample also includes a subsample of 324~MCs with \CeiO\ emission resolved.
Figure~\ref{f:map} shows the well-resolved MC sample on the \twCO\ integrated intensity and Galactic longitude-velocity maps.

\section{Results} \label{sec:result}
\subsection{Observational, morphological, and derived physical parameters} \label{sec:para}
We estimate the observational, morphological, and physical parameters for each MC in preparation for examining the properties of the MCs. 
The observational parameters of the \twCO, \thCO, and \CeiO~(J=1--0) line emissions for the MCs are as follows:
\begin{itemize}
\item {\it The Galactic longitude} ($L$, \degree): the intensity-weighted mean Galactic longitude by the \twCO~(J=1--0) line.
\item {\it The Galactic latitude} ($B$, \degree): the intensity-weighted mean Galactic latitude by the \twCO~(J=1--0) line.
\item {\it Systemic velocity} ($V_{\rm sys}$, \km\ps): the intensity-weighted mean LSR velocity of the \thCO~(J=1--0) emission. If no significant \thCO~(J=1--0) emission is detected, the velocity of the \twCO~(J=1--0) emission is applied.
\item {\it Spans in the Galactic longitude and latitude} ($L_{\rm span}$ and $B_{\rm span}$, \degree): the spans in the Galactic longitude and latitude directions for the \twCO~(J=1--0) line emission.
\item {\it Velocity span} ($V_{\rm span}$, \km\ps): the span in the velocity direction for the \twCO~(J=1--0) line emission.
\item {\it Velocity dispersion} ($\sigma_{\rm v}$, \km\ps): the square root of the intensity-weighted velocity variance (moment II). 
\item {\it Average integrated intensity} ($\overline{W}$, \K\km\ps): the average integrated intensity. 
\item {\it Peak temperature} ($T_{\rm peak}$, \K): the maximum brightness temperature within the cloud boundary.
\item {\it Peak temperature of the average spectrum} ($\overline{T}_{\rm peak}$, \K): the maximum brightness temperature of the average spectrum.
\item {\it Peak integrated intensity} ($W_{\rm peak}$, \K\km\ps): the maximum integrated intensity of the emission map.
\item {\it Area} ($A$, \amin$^2$): the angular area of the cloud.
\item {\it Flux} ($F$, \K\km\ps\amin$^2$): the overall flux of the cloud, which can be derived from the above quantities as $F=\overline{W} A$. 
\end{itemize}

The morphological parameters of the \twCO~(J=1--0) integrated intensity map for MCs are as follows:
\begin{itemize}
\item {\it Angular lengths of the principle axes} ($l_{\rm min,\,ang}$ and $l_{\rm maj,\,ang}$, \degree): FWHM along the minor and major principle axes, respectively, which are corrected for finite angular resolution.
\item {\it Oblique angle} ($\theta$, \degree): the angle between the long principle axis and the Galactic longitude axis.
\item {\it Ellipticity} ($e$): derived from the lengths of the minor and major axes as $1-l_{\rm min,\,ang}/l_{\rm maj,\,ang}$.
\item {\it RJ-plots parameters} ($R_1$ and $R_2$): $I_1$ and $I_2$ are the principle moments of inertia , with $I_1$ selected as the smaller one. If the cloud is perfectly circular, $I_1=I_2=(\overline{W} A^2)/(4 \pi)$, where A is the total map area and $\overline{W}$ is the average integrated intensity. $I_0$ is defined as $(\overline{W} A^2)/(4 \pi)$. The J-plots parameters are then defined as $J_i=(I_0-I_i)/(I_0+I_i)$, where i=1, 2, which can be used together as a reference to distinguish some cloud morphologies. The RJ-plots parameters are defined as $R_1=J_1-J_2$ and $R_2=J_1+J_2$ \citep[see][for details]{Jaffa+2018, Clarke+2022}. $R_1$ is proportional to $I_0(I_2-I_1)$ and is positive, which is a measure of MC symmetry. A closer value to zero indicates a more symmetrical morphology. The $R_1$ value is equivalent to the ellipticity $e$. The $R_2$ value is proportional to $I_0^2-I_1 I_2$, which can be used to measure the uniformity of the MC. Positive values imply a centrally enhanced density distribution, and negative values imply the opposite.
We note that these parameters are not suitable for distinguishing between shell-like from cross/hub-like structures.
\end{itemize}

The derived physical parameters of MCs are as follows:
\begin{itemize}
\item {\it Kinematic distance} ($D$, \kpc): the MC's kinematic distance, estimated based on its systemic velocity. The kinematic distance is derived from the Galactic rotation curve of \cite{BrandBlitz1993} with the Sun's Galactocentric distance of 8.15~\kpc\ and circular rotation speed of 236~\km\ps\ \citep{Reid+2019}. For MCs with positive systemic velocities in the G50 region, with non-unique kinematic distances, a full distance-probability density function method \citep{Reid+2016, Reid+2019} is applied to calculate the distance instead. For these MCs, we additionally eliminate the distances of MCs largely affected by near- and far-distance ambiguities. This includes MCs with masses exceeding the 1$\sigma$ range of the Larson $\sigma_{\rm v}$-$M$ relation fitted here and MCs with heights exceeding the 5$\sigma$ thickness of the Galactic thick molecular disk \citep[$\sim$590~pc;][]{Su+2021}. 
We also eliminate the distances of MCs that are less than the radius of the Local Bubble \citep[165~pc;][]{Zucker+2022}, since their velocities are dominated by proper motions rather than the Galactic rotation.
The distance estimation method is the same as that used in \citep{Zhou+2025}. 
496~MCs in the catalog have their distances eliminated and are therefore excluded from the analysis of derived physical parameters.
Additionally, MCs possibly located within the G120 spiral shock region may deviate from the Galactic rotation curve \citep[with systemic velocities ranging from $\sim$$-55$\km\ps\ to $\sim$$-30$\km\ps;][]{Zhou+2025}, so they are separated from the rest of the sample and presented independently. The catalog contains 336 such MCs.

The kinematic distances adopted in this study are based on the standard axisymmetric circular rotation model. It is well recognized that kinematic distances in the inner Milky Way are subject to significant systematic uncertainties due to variations in the rotation curve \citep{Sofue2023} and bar-driven non-circular streaming motions \citep{Baba2026a, Baba2026b}. While these effects are significantly mitigated for our sample, as all well-resolved MCs in our catalog have Galactocentric distances $R_{\rm gal} > 5$~kpc, they are located outside the bar-dominated region where the circular rotation approximation is valid \citep[e.g.,][]{Baba2026a}.
Residual uncertainties in our distance estimates primarily arise from the stochastic motion of MCs rather than from systematic shifts. The velocity dispersion is less than $\sim$11~\km\ps\ and decreases as the Galactocentric distance increases \citep{Zhou+2025}. 
Although such distance uncertainty can cause the distribution of derived physical parameters to broaden, it does not alter their characteristic values.
\item {\it Galactocentric distance} ($R_{\rm gal}$, \kpc): the distance to the Galactic center. It is derived as $R_{\rm gal}=\sqrt{D^2 {\rm cos}^2(B)+R_{\rm 0, gal}^2-2 D R_{\rm 0, gal}{\rm cos}(B) {\rm cos}(L)}$, where $R_{\rm 0,\,gal}=8.15$~\kpc is the Sun's distance to the Galactic center \citep{Reid+2019}.
\item {\it Galactic height} ($z_{\rm gal}$, \parsec): the height from the B=0\degree\ Galactic plane. It is derived as $z_{\rm gal}=D {\rm sin}(B)$.
\item {\it Lengths of principle axes} ($l_{\rm min}$ and $l_{\rm maj}$, \parsec): cloud sizes correspond to the angular lengths of the principle axes, $l_{\rm min,\,ang}$ and $l_{\rm maj,\,ang}$, which are derived as $l_{\rm min}=(l_{\rm min,\,ang}/1~{\rm radian}) D$ and $l_{\rm maj}=(l_{\rm maj,\,ang}/1~{\rm radian}) D$.
\item {\it Excitation temperature and optical depth} ($T_{\rm ex}$, \K) and $\tau$: they are estimated based on the average spectrum, under the assumption of local thermodynamic equilibrium (LTE) conditions. The calculation method is described in detail in \cite{Zhou+2016}. In the calculation, the area beam-filling factor of the \twCO~(J=1--0) emission $f$(\twCO) is assumed to be 1, and the area beam-filling factors of the \thCO\ and \CeiO~(J=1--0) emission, $f$(\thCO) and $f$(\CeiO), are calculated by their respective emission area ratios to the \twCO~(J=1--0) emission area. 
The optical depth of the \twCO~(J=1--0) emission is estimated as $\tau(^{12}{\rm CO})=\tau(^{13}{\rm CO}) f(^{13}{\rm CO}) [^{12}{\rm C}/^{13}{\rm C}]$, where the $[^{12}{\rm C}/^{13}{\rm C}]$ isotope ratio is estimated based on the Galactocentric distance of the cloud \citep{SunY+2024}.
The optical depth of the \CeiO~(J=1--0) emission is estimated as:
\begin{equation}
 \tau({\rm C^{18}O})=-{\rm log}(1-\frac{\overline{T}_{\rm peak}({\rm C^{18}O})/f({\rm C^{18}O})}{5.27(\frac{1}{{\rm exp}(5.27/T_{\rm ex})}-1)-0.166}).
\end{equation}
\item {\it CO column density} ($N$(CO), \cm$^{-2}$): \thCO\ and \CeiO\ column densities are estimated based on their optical depths. The column density of \twCO~(J=1--0) emission is estimated as $N(^{12}{\rm CO})=N(^{13}{\rm CO}) f(^{13}{\rm CO}) [^{12}{\rm C}/^{13}{\rm C}]$. The column density of \CeiO~(J=1--0) is estimated as:
\begin{equation}
N({\rm C^{18}O})=2.49\E{14} \frac{T_{\rm ex}}{1-{\rm exp}(-5.27/T_{\rm ex})} \tau({\rm C^{18}O}) \Delta v({\rm C^{18}O}),
\end{equation}
where $\Delta v({\rm C^{18}O})$ is the velocity width as FWHM for \CeiO\ calculated as $\Delta v({\rm C^{18}O})=2.355 \sigma_{\rm v}({\rm C^{18}O})$.
\item {\it {\rm H}$_2$ column density} ($N({\rm H}_2$), \cm$^{-2}$): $N({\rm H_2})_{13}$ and $N({\rm H_2})_{18}$ are derived from \thCO\ and \CeiO\ column densities by assuming their abundances of $1.4\E{-6}$ and $1.43\E{-7}$, respectively \citep{Ripple+2013}. $N({\rm H_2})_{12}$ is estimated using the conversion factor $N({\rm H_2})/W(^{12}{\rm CO})=1.8\E{20} \cm^{-2}\K^{-1}\km^{-1}\s$ \citep{Dame+2001}. 
\item {\it {\rm H}$_2$ number density} ($n({\rm H}_2$), \cm$^{-3}$): estimated as $n({\rm H}_2)=N({\rm H_2})/l_{\rm min}$ by assuming that the length of the MC in the line of sight (LoS) equals its minor principle axis length.
\item {\it Mass} ($M$, $\Msun$): estimated as $M=N({\rm H_2}) A D^2 \mu({\rm H_2}) m_{\rm H}$, where $\mu({\rm H_2})=2.8$ is the average molecular weight per hydrogen molecule \citep{Kauffmann+2008}, and $m_{\rm H}$ is the hydrogen atom mass.	
\item {\it Virial parameter} ($\alpha_{\rm vir}$): the virial mass is estimated as $M_{\rm vir}= k\,(\frac{\Delta v}{1~\km\s^{-1}})^2 l_{\rm maj}~\Msun$, where $k=k_0 e/{\rm arcsin}(e)$, $k_0=104.8$ \citep{MacLaren+1988}, $e$ is the eccentricity of the cloud, and $\Delta v$ is the velocity width as FWHM calculated as $\Delta v=2.355 \sigma_{\rm v}$. 	
The virial parameter is calculated as $\alpha_{\rm vir}=M_{\rm vir}/M$.
\item {\it Surface density} ($\Sigma$, $\Msun\parsec^{-2}$): calculated as $\Sigma=M/(A D^2)$. 
\item {\it Free-fall collapse time} ($\tau_{\rm ff}$, \yr): calculated as $\tau_{\rm ff}=\sqrt{(3\pi)/(32G\rho)}$, where G is the gravitational constant, $\rho=n({\rm H}_2) \mu({\rm H}_2) m_{\rm H}$ is the mass density.
\item{\it Thermal sound speed} ($c_{\rm s}$, \km\ps): derived as $c_{\rm s}=\sqrt{k_{\rm B} T_{\rm k}/\mu m_{\rm H}}$, where $k_{\rm B}$ is the Boltzmann constant, $T_{\rm k}$ is the kinematic temperature, which is assumed to be equal to the CO excitation temperature $T_{\rm ex}$, and $\mu=2.37$ is the average molecular weight per free particle \citep{Kauffmann+2008}.
\item{\it Non-thermal velocity dispersion} ($\sigma_{\rm v,\,nt}$, \km\ps): the thermal velocity dispersion in the LoS is estimated as $\sigma_{\rm v,\,th}=\frac{1}{3}\sqrt{k_{\rm B} T_{\rm k}/\mu_{\rm i} m_{\rm H}}$, where $\mu_{\rm i}$ is the molecular weight of the observed tracer. The non-thermal velocity dispersion in the LoS is estimated as $\sigma_{\rm v,\,nt}=\sqrt{\sigma_{\rm v}^2-\sigma_{\rm v,\,th}^2}$. We note that, if the velocity dispersion is isotropic, the total velocity dispersion can be estimated as three times the velocity dispersion in the LoS.
\item{\it Mach number} ($Mach$): derived as $Mach=3 \sigma_{\rm v,\,nt}/c_{\rm s}$.
\item{\it Total pressure} ($P_{\rm tot}/k_{\rm B}$, \K\cm$^{-3}$): derived as $P_{\rm tot}/k_{\rm B}=n({\rm H}_2) \mu({\rm H}_2) m_{\rm H} \sigma_{\rm v}^2/k_{\rm B}$.	
\item{\it Thermal pressure} ($P_{\rm thermal}/k_{\rm B}$, \K\cm$^{-3}$): derived as $P_{\rm thermal}/k_{\rm B}=n({\rm H}_2) T_{\rm ex}$.
\item{\it Turbulence time scale} ($\tau_{\rm turb}$, \yr): derived as $\tau_{\rm turb}=l_{\rm min}/\sigma_{\rm v,\,nt}$.
\item{\it Area ratio, or filling factor ($f$)}: the area ratios for \thCO\ and \CeiO\ are calculated as $f$(\thCO)$=A$(\thCO)$/A$(\twCO) and $f$(\CeiO)$=A$(\CeiO)$/A$(\twCO), respectively. 
\item{\it {\rm CO-to-H}$_2$ conversion factor} ($X_{\rm CO}$, $\cm^{-2}\K^{-1}\km^{-1}\s$): derived as $X_{\rm CO}=N({\rm H_2})_{13} f(^{13}{\rm CO})/\overline{W}$(\twCO).
\end{itemize}

\begin{center}
\begin{deluxetable}{cccccccc}
\tabletypesize{\scriptsize} 
\tablecaption{Observational and Morphological Parameters of MCs \label{tab:par0}}
\tablewidth{0pt}
\tablehead{
\colhead{Cloud ID\tablenotemark{a}}&\colhead{$l$}&\colhead{$b$}&\colhead{$V_{\rm sys}$}&\colhead{$l_{\rm min,\,ang}$(\twCO)}&\colhead{$l_{\rm maj,\,ang}$(\twCO)}&\colhead{$\sigma_{v}$(\twCO)}&\colhead{$\overline{W}$(\twCO)}\\
\colhead{}&\colhead{(deg)}&\colhead{(deg)}&\colhead{(\km\ps)}&\colhead{(deg)}&\colhead{(deg)}&\colhead{(\km\ps)}&\colhead{(K$ $\km\ps)}
}
\startdata
G126.916+00.361& 126.916& 0.361&$-$58.26&0.053&0.070&0.90&2.77\\
G127.395+01.157& 127.395& 1.157&$-$58.64&0.028&0.035&0.89&2.61\\
G121.966+02.004& 121.966& 2.004&$-$59.49&0.017&0.039&0.55&1.46\\
G128.616+02.792& 128.616& 2.792&$-$58.90&0.013&0.031&0.75&2.38\\
G126.208$-$00.220& 126.208& $-$0.220&$-$59.16&0.014&0.023&0.69\\
G125.991$-$00.197& 125.991& $-$0.197&$-$58.18&0.047&0.102&0.79\\
G123.814+01.602& 123.814& 1.602&$-$57.68&0.025&0.054&0.97&2.54\\
G122.220+02.156& 122.220& 2.156&$-$59.13&0.019&0.073&0.62&1.64\\
G123.466+02.273& 123.466& 2.273&$-$57.40&0.040&0.11&1.13&2.90\\
G127.031$-$00.153& 127.031& $-$0.153&$-$58.42&0.037&0.18&0.75\\
...\\

\enddata
\tablecomments{This table is available in its entirety in machine-readable form online.} 
\tablenotetext{a}{All cloud IDs share the prefix 'MWISP', which is omitted in this column.}
\end{deluxetable}
\end{center}
\begin{center}
\begin{deluxetable}{cccccccccc}
\tabletypesize{\scriptsize} 
\tablecaption{Derived Physical Parameters of MCs \label{tab:phy0}}
\tablewidth{0pt}
\tablehead{
\colhead{Cloud ID\tablenotemark{a}}&\colhead{$D$}&\colhead{$R_{\rm gal}$}&\colhead{$z_{\rm gal}$}&\colhead{$l_{\rm min}$(\twCO)}&\colhead{$l_{\rm maj}$(\twCO)}&\colhead{$T_{\rm ex}$}&\colhead{$N({\rm H}_2)_{12}$}&\colhead{$n({\rm H}_2)_{12}$}&\colhead{$M_{12}$}\\
\colhead{}&\colhead{(kpc)}&\colhead{(kpc)}&\colhead{(pc)}&\colhead{(pc)}&\colhead{(pc)}&\colhead{(K)}&\colhead{(cm$^{-2}$)}&\colhead{(cm$^{-3}$)}&\colhead{($\Msun$)}
}
\startdata
G126.916+00.361& 5.3& 12.1417& 33.7114&4.91&6.51& 4.4853&5.09$\times10^{20}$&32.89&1243.44\\
G127.395+01.157& 5.4& 12.2156& 109.281&2.63&3.29& 4.42204&4.86$\times10^{20}$&53.26&452.23\\
G121.966+02.004& 5.4& 11.9376& 189.487&1.62&3.66& 4.95637&2.72$\times10^{20}$&42.89&174.47\\
G128.616+02.792& 5.5& 12.3485& 267.861&1.20&2.97& 4.66274&4.45$\times10^{20}$&85.07&242.57\\
G126.208$-$00.220& 5.4& 12.1734& $-$20.8213&1.34&2.21& 4.93019&3.05$\times10^{20}$&62.38&115.20\\
G125.991$-$00.197& 5.3& 12.07& $-$18.2651&4.40&9.45& 4.3466&3.90$\times10^{20}$&27.27&733.21\\
G123.814+01.602& 5.2& 11.8817& 146.208&2.28&4.92& 4.38003&4.74$\times10^{20}$&61.17&443.20\\
G122.220+02.156& 5.4& 11.9184& 202.43&1.76&6.90& 4.56436&3.06$\times10^{20}$&56.66&256.50\\
G123.466+02.273& 5.2& 11.8374& 206.241&3.67&9.91& 4.05656&5.41$\times10^{20}$&44.54&978.84\\
G127.031$-$00.153& 5.4& 12.1645& $-$14.3342&3.46&16.46& 5.9627&7.02$\times10^{20}$&64.93&2776.36\\
...\\

\enddata
\tablecomments{This table is available in its entirety in machine-readable form online.} 
\tablenotetext{a}{All cloud IDs share the prefix 'MWISP', which is omitted in this column.}
\end{deluxetable}
\end{center}

The length of the MC along the LoS is assumed to be its minor principle axis length in the sky plane. Due to the projection effect, this measurement is uncertain for individual MCs. However, the uncertainty can be overcome when considering a large MC sample. 
The standard Galactic CO-to-H$_2$ conversion factor, $N({\rm H_2})/W(^{12}{\rm CO})=1.8\E{20} \cm^{-2}\K^{-1}\km^{-1}\s$ \citep{Dame+2001}, is used to estimate the total H$_2$ column densities $N({\rm H}_2)_{12}$ and other related parameters of MCs based on \twCO\ emission. 
The conversion factor can vary significantly on small scales within an MC 
\citep[e.g.,][]{SofueKohno2020, KohnoSofue2024}, therefore, this method is only valid when considering the cloud as a whole. Our study focuses on the statistical properties of MCs on a cloud-by-cloud basis. Nevertheless, the conversion factor can vary among different MCs depending on their physical conditions, such as metallicity \citep{Bolatto+2013}. Therefore, we also estimate this variation based on the H$_2$ column density derived from \thCO\ emission, $N({\rm H}_2)_{13}$. The conversion factor $X_{\rm CO}$ is then derived for each MC.
Tables~\ref{tab:par0} and \ref{tab:phy0} show the observational and derived physical parameters of individual MCs.

\subsection{Parameter distribution} \label{sec:paradist}
This section presents a systematic statistical analysis of the observational and physical properties of MCs based on the catalog constructed from the MWISP survey data.
First, we examine the overall distributions of various parameters. Second, we evaluate the impact of distance selection effects by comparing the overall sample with a nearby subsample to identify which parameter trends are intrinsic and which are observational biases. Third, we explore the distinct characteristics of MCs in specific environments, such as spiral shock regions and interarm spurs, as well as \CeiO-bright and supra-Larson velocity dispersion MC subsamples. Lastly, we synthesize a coherent picture of typical MC structure by comparing the physical states of molecular gas traced by the three CO isotopic lines. This highlights the transition from turbulence-dominated to gravity-dominated regimes. 

\subsubsection{Overview of Parameter Distributions}\label{sec:distover}

\begin{figure*}[ptbh!]
\centerline{{\hfil\hfil
\psfig{figure=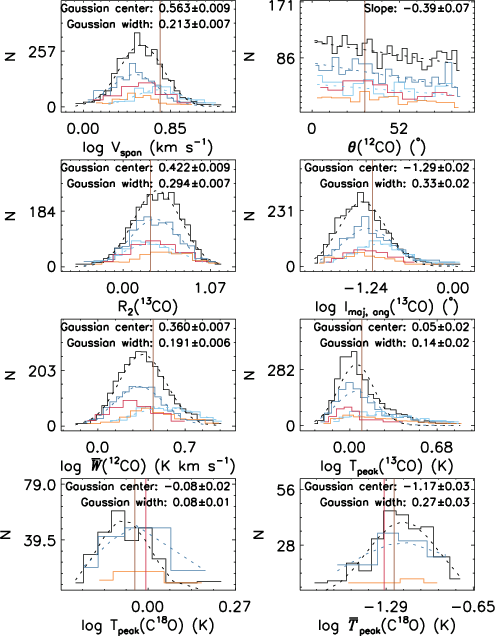,width=4.5in,angle=0, clip=}
\hfil\hfil}}
\caption{Distribution of observational parameters of identified MCs. We select eight observational parameters to show as examples. 
These parameters are $V_{\rm span}$, $\theta$(\twCO), $l_{\rm maj,\,ang}$(\twCO), $R_2$(\thCO), $l_{\rm min,\,ang}$(\thCO), $\overline{W}$(\twCO), $T_{\rm peak}$(\thCO), and $W_{\rm peak}$(\thCO).
The histograms for the overall MC sample without MCs possibly in the G120 spiral-shock region are shown by black solid lines and are fitted with a Gaussian function, except for the oblique angle $\theta$, which is fitted linearly. The fitting results are shown as black dotted lines.
The Gaussian parameters and the slope of the fitted line are shown in the upper-right corner of each panel.
For comparison, the histograms for MCs within the distance of 4~\kpc\ and with significant \CeiO\ emission are shown by blue and light blue solid lines, respectively. These lines are multiplied by a factor of two for better visibility.
These histograms are also fitted using the same methods as those used to fit the overall MC sample, and the corresponding fitting results are shown as blue and light blue dotted lines, respectively.
The red solid lines show the histograms of the supra-Larson velocity dispersion MCs (i.e., those with velocity dispersions greater than 1.5 times the fitted $\sigma_{\rm v}$-M relation).
The histograms of MCs that are possibly in the G120 spiral-shock region and in the G50 Perseus-Outer interarm spurs \citep[i.e., the red crosses in Figure~8 of][]{Zhou+2025} are shown by the orange and brown solid lines, respectively. 
When there are insufficient MCs, only the mean value of the parameters is shown as a vertical line instead of a histogram. 
All parameter distributions and the corresponding fitted parameters are available in the Science Data Bank at doi:10.57760/sciencedb.32354 \citep{Zhou+2026data}. 
}
\label{f:cldistpar}
\end{figure*}

\begin{figure*}[ptbh!]
\centerline{{\hfil\hfil
\psfig{figure=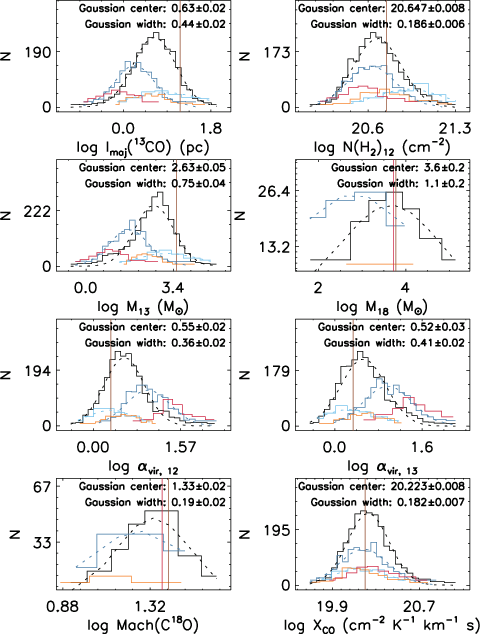,width=4.5in,angle=0, clip=}
\hfil\hfil}}
\caption{The same as Figure~\ref{f:cldistpar}, but for derived physical parameters. We select eight derived physical parameters to show as examples. 
These parameters are $l_{\rm min}$(\twCO), $l_{\rm min}$(\CeiO), $l_{\rm maj}$(\CeiO), $\sigma_{\rm v}$(\CeiO), $M_{13}$, $M_{18}$, $Mach$(\CeiO), and $X_{\rm CO}$.
All parameter distributions and the corresponding fitted parameters are available in the Science Data Bank at doi:10.57760/sciencedb.32354 \citep{Zhou+2026data}. 
}
\label{f:cldist}
\end{figure*}

The distributions of key observational and derived physical parameters for the identified MCs are presented in Figures~\ref{f:cldistpar} and \ref{f:cldist}, respectively. The full set of distributions and fitted parameters are available in the Science Data Bank at doi:10.57760/sciencedb.32354 \citep{Zhou+2026data}. 
Several distinct subsamples are compared, including the overall MC sample, a nearby MC sample ($D\le4\,kpc$), a \CeiO-bright MC sample (with resolved \CeiO~(J=1--0) emission), a supra-Larson velocity dispersion MC sample (with $\sigma_{\rm v}$ greater than 1.5 times the Larson $\sigma_{\rm v}$-M relation fitted here), a sample of MCs possibly within the G120 spiral-shock region, and a sample of MCs located in the G50 Perseus-Outer interarm spurs \citep[see the red crosses in Figure~8 of][]{Zhou+2025}.
We note that the overall sample excludes MCs possibly located in the G120 spiral-shock region, which is maintained throughout the following text. 

All of the observational and derived physical parameters are close to a log-normal distribution, except for the morphological parameters ($R_1$, $R_2$, and $e$), the oblique angle $\theta$, the distance $D$, the Galactocentric distance $R_{\rm gal}$, and the Galactic height $z_{\rm gal}$.
The morphological parameters $R_1$, $R_2$, and $e$, exhibit a normal distribution. 
The typical \twCO\ ellipticity $e$(\twCO) is 0.46 for the overall sample, indicating that MCs are generally oblate.
The oblique angle $\theta$ of MCs shows a nearly uniform distribution, thought there is a tendency for it to align parallel to the Galactic disk.
These imply that the morphology of MCs is predominantly governed by a dominant physical process and is influenced by the gravitational potential of the Galactic disk. The \CeiO\ oblique angle distribution is significantly flatter than those of \twCO\ and \thCO, because the shape of \CeiO-bright molecular gas is more dominated by self-gravity than by the external gravitational potential of the Galactic disk.

Several other parameters also have special distributions. The distance distribution of MCs has several peaks, suggesting an accumulation along spiral arms, which is also visible in the $R_{\rm gal}$ distribution. 
For the angular area of \CeiO\ emission, as its value decreases, the number of MCs increases continuously, implying a large population of undetected small-scale \CeiO\ emission regions within MCs.

The compensated \twCO\ average integrated intensity ($\overline{W}_{\rm all}$) is also examined, which corrects for missing flux below $3\sigma$ by assuming a Gaussian tail in the average spectrum. $\overline{W}_{\rm all}$ is slightly larger than the average integrated intensity $\overline{W}$, with the difference being more pronounced for small MCs. Nevertheless, the widths of their distributions are consistent. 

The CO-to-H$_2$ conversion factor, $X_{\rm CO}$, follows a log-normal distribution. While its mean value varies for different subsamples, the differences are all small and fall entirely within the standard deviation. 
The conversion factor used to derive the \twCO-related parameters \citep[$1.8\E{20} \cm^{-2}\K^{-1}\km^{-1}\s$;][]{Dame+2001} is consistent with the $X_{\rm CO}$ value estimated here.
The uncertainty in the conversion factor is suggested to be $\sim$30\% \citep{Bolatto+2013}, which is also approximately consistent with the dispersion of $X_{\rm CO}$ in this study. 
As the conversion factor and its related parameters follow log-normal distributions, their logarithms are used for the distribution and correlation analyses. After logarithmic transformation, the impact of uncertainty in the conversion factor is reduced. In fact, the uncertainty in the conversion factor is smaller than the intrinsic variation of the cloud properties.
Therefore, while the uncertainty in the conversion factor introduces some scatter, it does not significantly alter the statistical results.

\begin{center}
\begin{deluxetable}{cccccccc}
\tabletypesize{\scriptsize} 
\tablecaption{Typical Values of Observational and Morphological Parameters}\label{tab:par}
\tablewidth{0pt}
\tablehead{
\colhead{Obs. Para.}&\multicolumn{2}{c}{Overall MC sample}&\multicolumn{2}{c}{Nearby MC sample}&\multicolumn{3}{c}{\CeiO-bright MC sample}\\
&\colhead{\twCO}&\colhead{\thCO}&\colhead{\twCO}&\colhead{\thCO}&\colhead{\twCO}&\colhead{\thCO}&\colhead{\CeiO}
}
\startdata
log\,$l_{\rm min,\,ang}$ ($^{\circ}$)& $-$1.53 (0.29)& $-$1.60 (0.32)& $-$1.45 (0.31)& $-$1.52 (0.34)& $-$1.31 (0.35)& $-$1.37 (0.37)& $-$1.43 (0.38)  \\
log\,$l_{\rm maj,\,ang}$ ($^{\circ}$)& $-$1.23 (0.33)& $-$1.27 (0.35)& $-$1.13 (0.35)& $-$1.18 (0.37)& $-$0.98 (0.39)& $-$1.01 (0.40)& $-$1.02 (0.38)  \\
log\,$\sigma_{\rm v}$ (\km\ps)& $-$0.08 (0.11)& $-$0.10 (0.12)& $-$0.1 (0.1)& $-$0.14 (0.12)& $-$6.9$\E{-3}$ (0.12)& $-$0.03 (0.12)& $-$0.05 (0.15)  \\
log\,$\overline{W}$ (\K\km\ps)& 0.37 (0.20)& $-$0.63 (0.29)& 0.34 (0.20)& $-$0.71 (0.29)& 0.54 (0.22)& $-$0.38 (0.33)& $-$1.22 (0.27)  \\
log\,$T_{\rm peak}$ (\K)& 0.68 (0.15)& 0.09 (0.19)& 0.72 (0.15)& 0.10 (0.20)& 0.80 (0.17)& 0.27 (0.25)& $-$0.06 (0.1)  \\
log\,$W_{\rm peak}$ (\K\km\ps)& 0.88 (0.26)& 0.09 (0.31)& 0.86 (0.25)& 0.05 (0.30)& 1.13 (0.30)& 0.41 (0.37)& $-$0.12 (0.21)  \\
log\,$\overline{T}_{\rm peak}$ (\K)& 0.15 (0.14)& $-$0.74 (0.23)& 0.19 (0.15)& $-$0.76 (0.24)& 0.19 (0.16)& $-$0.65 (0.28)& $-$1.20 (0.25)  \\
log\,Area (\amin$^2$)& 1.35 (0.47)& 0.76 (0.53)& 1.49 (0.52)& 0.82 (0.59)& 1.74 (0.59)& 1.30 (0.66)& 0.38 (0.54)  \\
log\,Flux (\K\km\ps\amin$^2$)& 1.72 (0.57)& 0.72 (0.59)& 1.83 (0.63)& 0.78 (0.64)& 2.29 (0.70)& 1.37 (0.72)& 0.53 (0.54)  \\
$R_1$& 0.61 (0.31)& 0.63 (0.32)& 0.64 (0.32)& 0.65 (0.34)& 0.68 (0.33)& 0.70 (0.34)& 0.72 (0.35)  \\
$R_2$& 0.24 (0.23)& 0.42 (0.30)& 0.21 (0.22)& 0.39 (0.31)& 0.18 (0.26)& 0.32 (0.31)& 0.37 (0.37)  \\
$\theta_{\rm oblique}$ ($^{\circ}$)& 42 (27)& 42 (27)& 41 (26)& 41 (26)& 40 (27)& 41 (27)& 41 (27)  \\
$e$& 0.46 (0.19)& 0.48 (0.19)& 0.47 (0.19)& 0.49 (0.19)& 0.49 (0.19)& 0.51 (0.19)& 0.54 (0.20)  \\

\enddata
\tablecomments{The overall MC sample does not include MCs that are possibly in the G120 spiral-shock region. The mean values of each parameter are shown, and the corresponding variance values are enclosed in parentheses.}
\end{deluxetable}
\end{center}

\begin{center}
\begin{deluxetable}{cccccccc}
\tabletypesize{\scriptsize} 
\tablecaption{Typical Values of Derived Physical Parameters \label{tab:phy}}
\tablewidth{0pt}
\tablehead{
\colhead{Phy. Para.}&\multicolumn{2}{c}{Overall MC sample}&\multicolumn{2}{c}{Nearby MC sample}&\multicolumn{3}{c}{\CeiO-bright MC sample}\\
&\colhead{\twCO}&\colhead{\thCO}&\colhead{\twCO}&\colhead{\thCO}&\colhead{\twCO}&\colhead{\thCO}&\colhead{\CeiO}
}
\startdata
log\,$l_{\rm min}$ (\parsec)& 0.36 (0.42)& 0.29 (0.44)& $-$0.04 (0.39)& $-$0.11 (0.41)& 0.58 (0.45)& 0.52 (0.47)& 0.48 (0.48)  \\
log\,$l_{\rm maj}$ (\parsec)& 0.66 (0.44)& 0.62 (0.46)& 0.28 (0.41)& 0.23 (0.43)& 0.90 (0.50)& 0.86 (0.52)& 0.89 (0.47)  \\
log\,$T_{\rm ex}$\tablenotemark{a} (\K)& 0.66 (0.06)& -& 0.67 (0.06)& -& 0.70 (0.07)& -& - \\
log\,$\tau$& 0.99 (0.18)& $-$0.15 (0.25)& 0.94 (0.18)& $-$0.11 (0.27)& 1.04 (0.18)& $-$0.32 (0.16)& 0.02 (0.31)  \\
log\,$N$(CO) (\cm$^{-2}$)& 16.48 (0.19)& 15.34 (0.19)& 16.40 (0.18)& 15.35 (0.21)& 16.67 (0.23)& 15.30 (0.15)& 15.62 (0.28)  \\
log\,$N({\rm H}_2)$ (\cm$^{-2}$)& 20.65 (0.20)& 21.19 (0.19)& 20.61 (0.19)& 21.21 (0.21)& 20.89 (0.21)& 21.16 (0.15)& 22.47 (0.28)  \\
log\,$n({\rm H}_2)$ (\cm$^{-3}$)& 1.76 (0.38)& 2.35 (0.48)& 2.13 (0.31)& 2.78 (0.47)& 1.79 (0.40)& 2.11 (0.49)& 3.47 (0.53)  \\
log\,$M$ ($\Msun$)& 2.58 (0.89)& 2.55 (0.86)& 1.72 (0.85)& 1.65 (0.79)& 3.2 (1.0)& 3.17 (0.88)& 3.61 (0.83)  \\
log\,$\alpha_{\rm vir}$& 0.61 (0.42)& 0.55 (0.45)& 1.01 (0.41)& 0.98 (0.42)& 0.31 (0.35)& 0.31 (0.39)& $-$0.16 (0.40)  \\
log\,$\Sigma$ ($\Msun\parsec^{-2}$)& 1.00 (0.20)& 1.54 (0.19)& 0.96 (0.19)& 1.56 (0.21)& 1.24 (0.21)& 1.51 (0.15)& 2.82 (0.28)  \\
log\,$\tau_{\rm ff}$ (\yr)& 6.61 (0.19)& 6.31 (0.24)& 6.42 (0.16)& 6.10 (0.24)& 6.59 (0.20)& 6.43 (0.25)& 5.75 (0.27)  \\
log\,$c_{\rm s}$\tablenotemark{a} (\km\ps)& $-$0.90 (0.03)& -& $-$0.89 (0.03)& -& $-$0.88 (0.04)& -& - \\
log\,$\sigma_{\rm v,\,nt}$ (\km\ps)& $-$0.07 (0.11)& $-$0.10 (0.12)& $-$0.1 (0.1)& $-$0.14 (0.12)& $-$9.2$\E{-3}$ (0.12)& $-$0.03 (0.13)& $-$0.05 (0.15)  \\
log\,$Mach$& 1.30 (0.12)& 1.28 (0.13)& 1.26 (0.12)& 1.23 (0.13)& 1.35 (0.14)& 1.32 (0.15)& 1.32 (0.16)  \\
log\,$P_{\rm tot}/k_{\rm B}$ (\K\cm$^{-3}$)& 4.14 (0.38)& 4.69 (0.42)& 4.43 (0.31)& 5.02 (0.41)& 4.31 (0.38)& 4.58 (0.38)& 5.90 (0.52)  \\
log\,$P_{\rm thermal}/k_{\rm B}$ (\K\cm$^{-3}$)& 2.41 (0.40)& 3.01 (0.49)& 2.79 (0.32)& 3.45 (0.47)& 2.50 (0.44)& 2.81 (0.53)& 4.18 (0.55)  \\
log\,$\tau_{\rm turb}$ (\yr)& 6.42 (0.37)& 6.38 (0.38)& 6.06 (0.34)& 6.02 (0.36)& 6.58 (0.37)& 6.54 (0.39)& 6.52 (0.42)  \\
log\,$f$\tablenotemark{b}& -& $-$0.58 (0.22)& -& $-$0.66 (0.25)& -& $-$0.35 (0.17)& $-$1.22 (0.23)  \\
log\,$X_{\rm CO}$\tablenotemark{a} (\cm$^{-2}$\K$^{-1}$\km$^{-1}$\s)& 20.24 (0.20)& -& 20.20 (0.22)& -& 20.18 (0.19)& -& - \\

\enddata
\tablecomments{The overall MC sample does not include MCs that are possibly in the G120 spiral-shock region. The mean values of each parameter are shown, with the corresponding variance values enclosed in parentheses.}
\tablenotetext{a}{Estimated based on both \twCO\ and \thCO~(J=1--0) spectra, assuming local thermodynamic equilibrium (LTE) conditions.}
\tablenotetext{b}{Area ratio of the corresponding lines to the \twCO~(J=1--0) line emission.}
\end{deluxetable}
\end{center}

Typical parameter values for the overall, nearby, and \CeiO-bright  MC samples are listed in Tables~\ref{tab:par} and \ref{tab:phy}.
F-tests and t-tests are performed to assess the statistical significance of differences in the mean and variance of parameters between the overall sample and other subsamples, including the nearby sample, \CeiO-bright sample, supra-Larson velocity dispersion sample, sample in the G120 spiral-shock region, and sample in the G50 Perseus-Outer interarm spurs. The comparison results are presented in Sections~\ref{sec:disteff}, \ref{sec:subsampprop}, and \ref{sec:tracercomp}.

\subsubsection{Effect of Distance on Parameter Distributions} \label{sec:disteff}
Due to observational limitations, only larger MCs can be resolved at greater distances, which biases parameter distributions. This issue is common to all survey observations. To evaluate distance effects, we constructed a nearby MC sample ($D\le4\kpc$), within which MCs larger than 1 pc are resolved.

The nearby sample exhibits a smaller mean of the average integrated intensity $\overline{W}$ and a larger mean angular size among the observational and morphological parameters. 
These differences propagate to other related parameters. 
The detection of weaker MC emission at closer distances, leading to larger apparent MC extents, is likely due to a larger beam filling factor for the faint parts of nearby MCs. Similarly, the mean \twCO\ peak temperature $T_{\rm peak}$(\twCO) is marginally larger for the nearby sample, attributable to a larger beam filling factor for the bright parts. 
Although all identified MCs are spatially resolved, making the \twCO\ peak temperature of the average spectrum basically distance-independent, its mean is still larger for the nearby sample. This primarily stems from the properties of small MCs, indicating their higher excitation temperatures.
The means of the RJ-plots parameters $R_2$ and  the \thCO\ peak integrated intensity $W_{\rm peak}$(\thCO) of the nearby MC sample are smaller. This implies a less centrally peaked density distribution, a property intrinsic to small MCs that is not directly distance-dependent.
Velocity dispersion also depends on distance due to the Larson $\sigma_{\rm v}$-L relation \citep{Larson1981}.
The widths of the distributions of observational parameters are not significantly affected by distance, except for the \thCO\ angular length of the minor axis $l_{min\,ang}$(\thCO), \twCO\ velocity dispersion, and \twCO\ and \thCO\ fluxes $F$(\twCO, \thCO). The flux means do not change significantly.

The nearby MC sample contains a more complete population of small MCs, with a size distribution shifted toward smaller scales compared to the overall sample.
Consequently, this alters size-dependent physical parameters. 
For example, the mean mass $M$ is smaller for the nearby sample (see Figure~\ref{f:cldist} and Table~\ref{tab:phy}).
These differences highlight the properties of small MCs, including a higher excitation temperature $T_{\rm ex}$, higher H$_2$ number density $n$(H$_2$), larger virial parameter $\alpha_{\rm vir}$, smaller \thCO\ area ratio, and other related parameters.
The mean CO-to-H$_2$ conversion factor $X_{\rm CO}$ is also slightly smaller for small MCs. 
The oblique angle $\theta$ of small MCs is more uniformly distributed (slope $\sim$$-0.21$; see Figure~\ref{f:cldistpar}).
Due to the limitations of the Galactic latitude boundaries of the observations, the mean Galactic height ($z_{\rm gal}$) of the nearby sample is smaller than that of the overall sample. 
We note that the variance of the size does not change significantly with distance. 
However, both the mean and variance of several physical parameters show significant distance dependence. These parameters include Galactic height $z_{\rm gal}$, H$_2$ number density derived from \twCO\ $n({\rm H_2})_{12}$, free-fall collapse time derived from \twCO\ $\tau_{\rm ff,\,12}$, total pressure derived from \twCO\ $P_{\rm tot,\,12}/k_{\rm B}$, thermal pressure derived from \twCO\ $P_{\rm thermal,\,12}/k_{\rm B}$, \thCO\ area ratio $f$(\thCO), and CO-to-H$_2$ conversion factor $X_{\rm CO}$.
The distributions of these parameters are strongly distance-dependent and should be treated with caution in studies of MCs.

\subsubsection{Properties of Specific Subsamples} \label{sec:subsampprop}
The parameters of MCs possibly in the G120 Perseus arm spiral shock region and the G50 Perseus-Outer interarm spurs are also affected by distance (e.g., physical size).
The mean distances of these two subsamples are smaller and larger than those of the overall sample, respectively.
The parameters of these samples that exhibit significant differences relative to the overall sample, and where these differences are opposite to the general distance-induced trend, can be used to reveal their unique properties.
For the G120 region sample, the means of R$_2$(\thCO), $\overline{W}$(\twCO,\,\thCO), $W_{\rm peak}$(\twCO,\,\thCO), $N$(H$_2$)$_{12}$, and $f$(\thCO) are larger, while $e$(\thCO) and $\alpha_{\rm vir,\,12}$ are smaller. These differences cannot be explained by distance alone and suggest a significant amount of concentrated and elongated \thCO\ gas within these MCs. Their size and mass show small variations, while the excitation temperature varies significantly.
In the G50 interarm spurs, the number of MCs is small, and only a few parameters show significant differences. 
Most of these differences are distance-induced, except for relatively large $e$(\twCO,\,\thCO) and relatively small $\tau$(\thCO). This suggests that these MCs are more elongated. The variations of $l_{\rm min}$(\twCO) and $M_{13}$ for these MCs are also larger.

The mean distance of the \CeiO-bright sample is comparable to that of the overall sample, which minimizes the distance effects on parameter differences. As expected, \CeiO-bright MCs have larger angular and physical sizes. Consequently, their $T_{\rm peak}, \overline{W}, W_{\rm peak}, A, F, \sigma_{\rm v}, M$, and $Mach$ are relatively large.
Compared to the overall sample, these large MCs also have a larger ellipticity ($e$), indicating a more elongated shape. Higher excitation temperatures ($T_{\rm ex}$) and higher dense gas ratios ($f$) are also evident. Notably, the average $\tau$(\thCO) is relatively small, likely due to efficient self-shielding against dissociating photons in large MCs. Their $X_{\rm CO}$ is also slightly smaller.

Supra-Larson velocity dispersion MCs are predominantly nearby and mostly small. Unlike the general nearby sample, they have smaller means of $T_{\rm peak}$(\twCO,\,\thCO), $\overline{T}_{\rm peak}$(\twCO,\,\thCO), and $T_{\rm ex}$ compared to the overall sample. Their $A$(\thCO) is smaller, while their $\tau$(\thCO) and $X_{\rm CO}$ are larger. These MCs tend to exhibit small variations in properties such as size and mass. Their excess velocity dispersion, small size, low \thCO\ gas content, and low excitation temperature suggest that they are likely young MCs that have inherited perturbations from the more diffuse ISM.

\subsubsection{Comparison Across Molecular Tracers} \label{sec:tracercomp}
\begin{figure*}[ptbh!]
\centerline{{\hfil\hfil
\psfig{figure=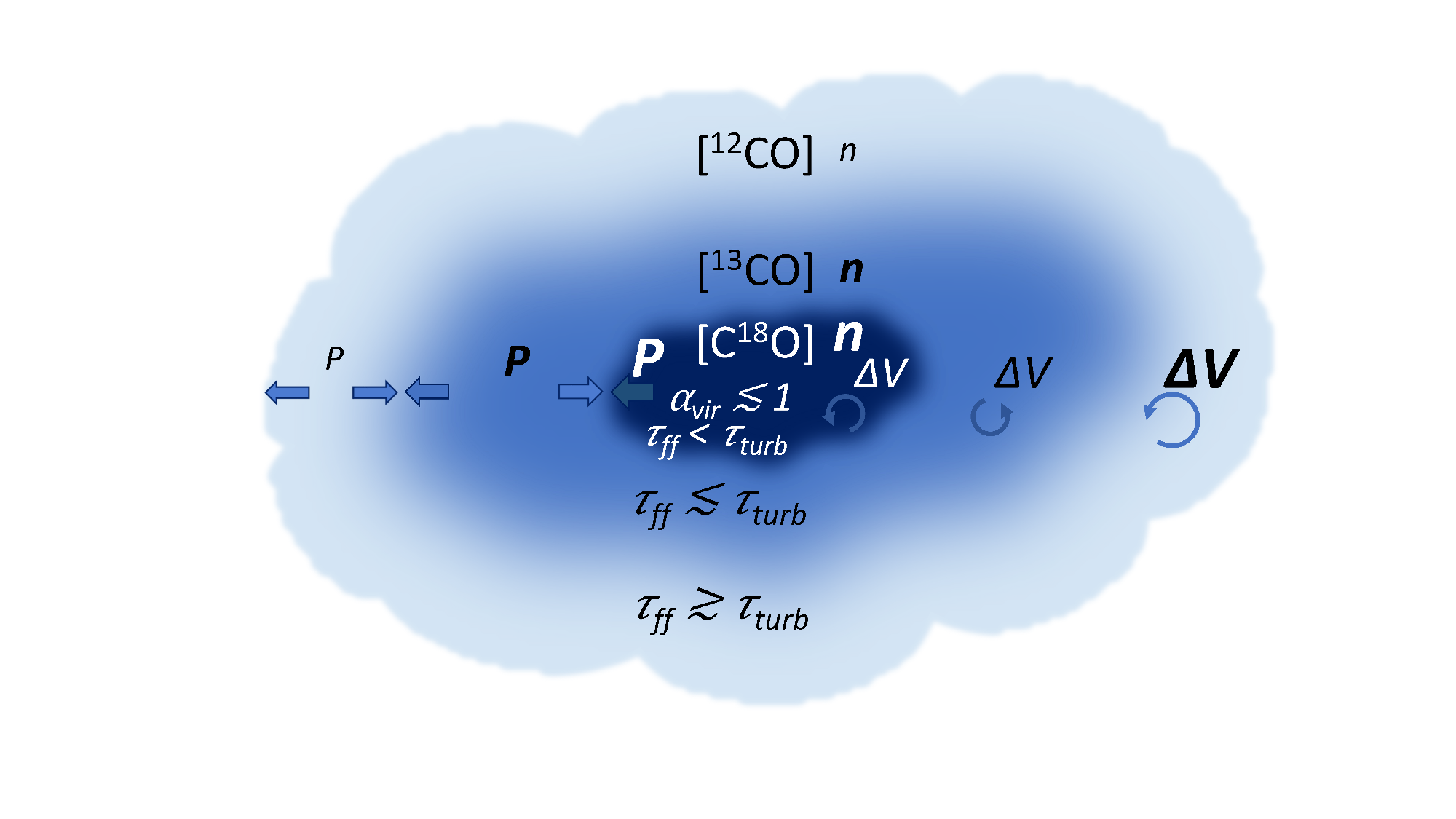,height=3.5in,angle=0, clip=}
\hfil\hfil}}
\caption{Schematic view of the MC structure, illustrated with typical parameter values estimated using \twCO, \thCO, and \CeiO~(J=1--0) lines. The size of each parameter sign indicates the magnitude of its typical value. Color depth indicates H$_2$ number density ($n$), thick arrows indicate pressure ($P$), and circular arrows indicate turbulence ($\Delta v$). The actual \thCO- and \CeiO-bright regions are smaller than shown in the schematic, and some of them are clumpy.
The typical area ratio of \thCO\ to \twCO\ is $\sim$0.26 for the overall MC sample. For MCs with significant \CeiO\ emission, the typical area ratio of \thCO\ to \twCO\ is $\sim$0.45, and the \CeiO\ to \twCO\ ratio is $\sim$0.06.
The stratified structure of \twCO\ and \thCO\ remains valid for MCs in which no \CeiO\ emission is detected.
}
\label{f:mcview}
\end{figure*}

The physical parameters derived from different lines are compared. 
The comparison results for \CeiO\ are based on the \CeiO-bright subsample.
The comparison results for \twCO\ and \thCO\ obtained from the overall sample and \CeiO-bright subsample are consistent, although the results from the larger overall sample are more significant.
The relative proportions of the gases traced by the lines can be examined by their respective area ratios.
The typical \thCO\ area ratio $f$(\thCO) is $\sim$0.26 for the overall MC sample. For the nearby sample containing a more complete set of small MCs, this ratio decreases to $\sim$0.22. 
\CeiO-bright MCs, being relatively large, have a larger typical \thCO\ area ratio of $\sim$0.45, with a corresponding \CeiO\ area ratio $f$(\CeiO) of $\sim$0.06. 
The average integrated intensity fractions for the overall MC sample are $\overline{W}$(\thCO)/$\overline{W}$(\twCO)$\sim$0.11 and $\overline{W}$(\CeiO)/$\overline{W}$(\twCO)$\sim$0.01. These values are lower than the corresponding area ratios, and are comparable to those detected in the FUGIN survey \citep[see Figure~12 in][]{Torii+2019}.


Turbulence (Mach number $Mach$) is stronger in the more diffuse \twCO\ gas than in the \thCO\ gas.
All measured velocity dispersions are supersonic. However, due to the turbulence cascade, velocity dispersion decreases to subsonic velocities at small scales. 
Using typical values for the principle axis length $l_{\rm maj}$ and Mach number, the transition scale from supersonic to subsonic turbulence is estimated to be 0.01~\parsec, according to the relation for the sonic scale \citep[$0.42 Mach^{-2.04} l_{\rm maj}$;][]{Federrath+2021}. 
This value is consistent with the sonic scales observed in high-resolution studies \citep{Sokolov+2018, Li+2020, Yue+2021}.
For H$_2$ number density $n({\rm H}_2)$, total pressure $P_{\rm tot}/k_{\rm B}$, and thermal pressure $P_{\rm thermal}/k_{\rm B}$, the trend reverses. \twCO-bright gas has the smallest typical value, and the \thCO\ value is smaller than the \CeiO\ value. This is consistent with the higher critical densities of the \thCO\ and \CeiO~(J=1--0) lines, which enable them to trace denser gas than the \twCO~(J=1--0) line. 
The inward-increasing pressure gradient can be balanced by the gravity of centrally denser MCs, atomic gas pressure, magnetic fields, etc. The means of the RJ-plot parameters $R_2$(\thCO,\,\CeiO) are larger than that for \twCO, indicating that the density distributions of \thCO\ and \CeiO\ gases are more centrally concentrated. 
In general, \thCO\ and \CeiO-bright gases are concentrated in the inner regions of MCs, while \twCO-bright gas is more diffuse, less dense, and more turbulent.

When comparing the free-fall collapse time ($\tau_{\rm ff}$) and turbulence timescale ($\tau_{\rm turb}$), we find that $\tau_{\rm ff}$ is longer than $\tau_{\rm turb}$ for \twCO-bright gas, but slightly shorter for \thCO-bright gas and much shorter for \CeiO-bright gas. This suggest that \thCO- and \CeiO-bright molecular gas has a sufficient lifetime to collapse, though not all \twCO-bright gas does. Molecular gas that is bright only in \twCO\ emission is typically perturbation-dominated and does not contribute directly to star formation. This property may extend to CO-dark molecular gas in the outer layers of MCs. \thCO-bright gas appears to be a turning point at which gravity becomes significant. This is consistent with the finding that dense gas plays a key role in star formation \citep[e.g., as demonstrated by the Gao-Solomon relation, etc.;][]{GaoSolomon2004, Lada+2010, Neumann+2024, Jiao+2025}. 
The mean virial parameter estimated from \CeiO\ is less than unity, indicating that gravity dominates over turbulence. Conversely, mean virial parameters from \twCO\ and \thCO\ are greater than unity. Figure~\ref{f:mcview} shows a schematic diagram illustrating the typical MC structure as revealed by these three isotopic lines.

While our statistical sample represents the general population of MCs, the formation of such structures is often influenced by large-scale galactic dynamics. For instance, the 'collect-and-collapse' scenario \citep[e.g.,][]{Deharveng+2010} suggests that expanding \HII\ regions can sweep up neutral material into dense shells, potentially contributing to the dense rims or outer layers observed in our \twCO-bright regions. Similarly, \cite{Inutsuka+2015} propose a sequential MC formation scenario driven by multiple overlapping expanding bubbles, which can explain the oblate morphology observed in our sample. In contrast, more extreme triggering mechanisms, such as cloud-cloud collisions (CCC), have been identified as key drivers for massive star formation in specific regions \citep[e.g.,][]{Fukui+2021}. CCCs typically produce high velocity dispersion and distinct kinematic signatures, such as velocity bridges. 
The supra-Larson velocity dispersion subsample identified in this study may represent clouds where such shock-compression events have recently occurred, whereas the majority of our sample reflects clouds for which the internal quasi-steady state may be better explained by a transient picture \citep[as studied in a series of works by][]{Yuan+2024, YuanYang2025}.

\subsection{Parameter correlation} \label{sec:paracorr}
\subsubsection{Correlations and Scaling Relations}\label{sec:corr}
\begin{figure*}[ptbh!]
\centerline{{\hfil\hfil
\psfig{figure=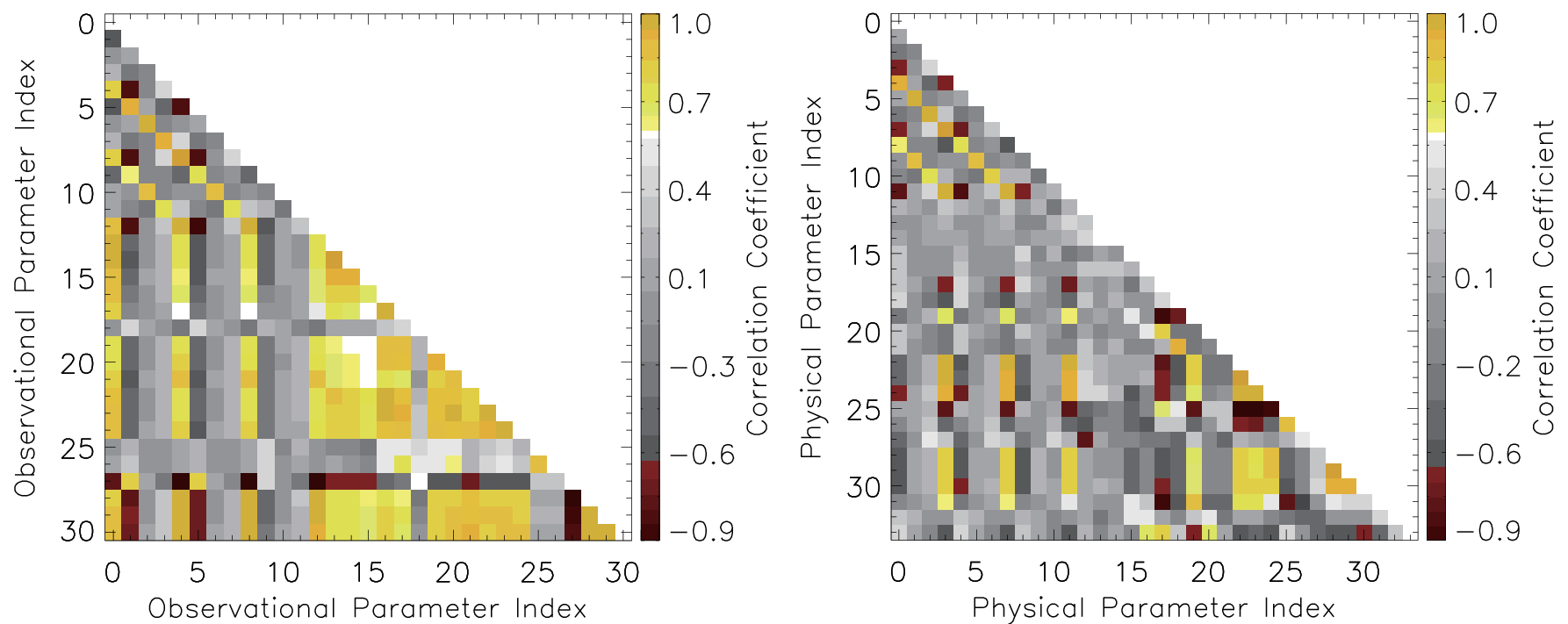,height=2.5in,angle=0, clip=}
\hfil\hfil}}
\caption{Correlation matrix of different observational ({\it left}) and derived physical ({\it right}) parameters of MCs.
The observational parameters corresponding to indexes 0 through 30 are 0: log~$V_{\rm span}$, (1: $R_2$, 2: $\theta$, 3: $e$, 4: log~$l_{\rm max,\,ang}$) for \twCO, (5: $R_2$, 6: $\theta$, 7: $e$, 8: log~$l_{\rm maj,\,ang}$) for \thCO, (9: $R_2$, 10: $\theta$, 11: $e$, 12: log~$l_{\rm maj,\,ang}$) for \CeiO, 13\&14\&15: log~$\sigma_{\rm v}$(\twCO,\,\thCO,\,\CeiO), 16\&17\&18: log~$\overline{W}$(\twCO,\,\thCO,\,\CeiO), 19\&20\&21: log~$T_{\rm peak}$(\twCO,\,\thCO,\,\CeiO), 22\&23\&24: log~$W_{\rm peak}$(\twCO,\,\thCO,\,\CeiO), 25\&26\&27: log~$\overline{T}_{\rm peak}$(\twCO,\,\thCO,\,\CeiO), and 28\&29\&30: log~$A$(\twCO,\,\thCO,\,\CeiO), respectively.
The physical parameters corresponding to indexes 0 through 36 are (0: $R_2$, 1: $\theta$, 2: $e$, 3: log~$l_{\rm maj}$) for \twCO, (4: $R_2$, 5: $\theta$, 6: $e$, 7: log~$l_{\rm maj}$) for \thCO, (8: $R_2$, 9: $\theta$, 10: $e$, 11: log~$l_{\rm maj}$) for \CeiO, 12\&13\&14: log~$\sigma_{\rm v}$(\twCO,\,\thCO,\,\CeiO), 15: $D$, 16: $R_{\rm gal}$, 17: log~$z_{\rm gal}$, 18: log~$T_{\rm ex}$, 19\&20\&21: log~$\tau$(\twCO,\,\thCO,\,\CeiO), 22\&23\&24: log~$N({\rm H_2})_{(12, 13, 18)}$, 25\&26\&27: log~$M_{12, 13, 18}$, 28: log~$\alpha_{\rm vir,\,12}$, 29: log~$\alpha_{\rm vir,\,13}$, 30: log~$\alpha_{\rm vir,\,18}$, 31\&32\&33: log~$Mach$(\twCO,\,\thCO,\,\CeiO), 34\&35: log~$f$(\thCO,\,\CeiO), and 36: log~$X_{\rm CO}$, respectively.
}
\label{f:matrix}
\end{figure*}

Figure~\ref{f:matrix} shows the correlation coefficients between different parameters. Most parameters exhibit correlations with others to some extent. We exclude parameters that are linear combinations of others from the correlation analysis, for example, flux $F$, the RJ-plot parameter $R_1$, the length of the minor principle axis $l_{\rm min}$, etc. 
The sample includes numerous large and small MCs, giving the correlation results broad applicability.

\begin{figure*}[ptbh!]
\centerline{{\hfil\hfil
\psfig{figure=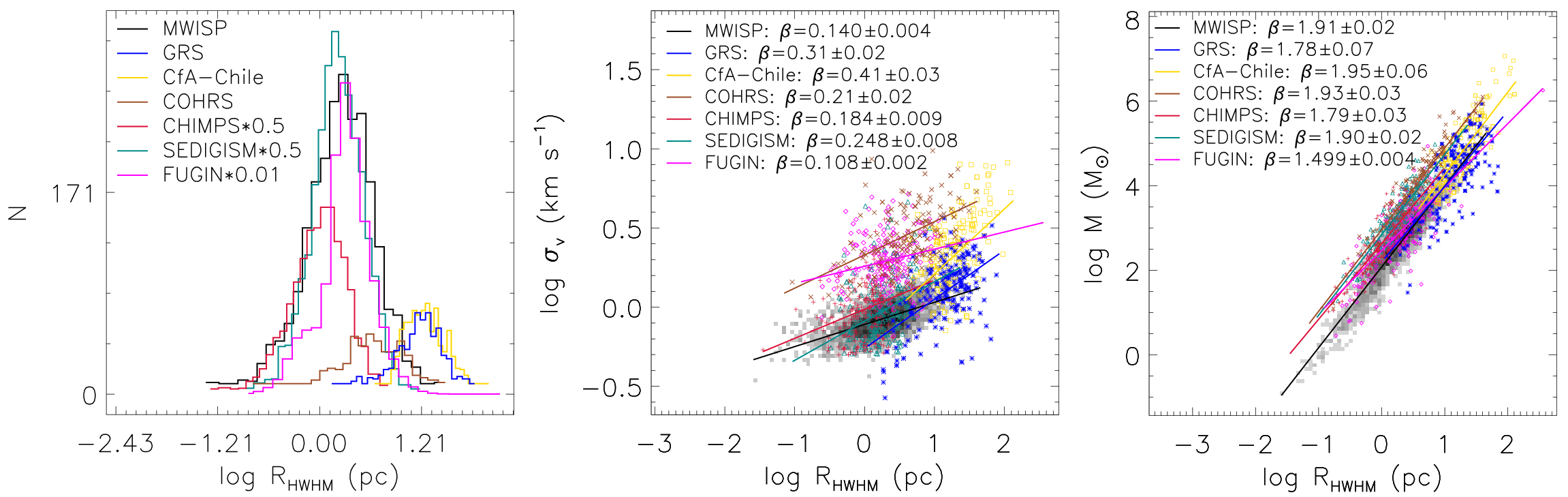,width=6.2in,angle=0, clip=}
\hfil\hfil}}
\caption{Histogram of radius (left), radius-velocity dispersion relation (middle), and radius–mass relation (right) for well-resolved MWISP clouds (\twCO, \thCO, and \CeiO~J=1--0), overlaid with those of well-resolved clouds from other literature catalogs. These include GRS \citep[\thCO~J=1--0;][]{Rathborne+2009, Roman-Duval+2009, Roman-Duval+2010}, CfA-Chile \citep[\twCO~J=1--0;][]{Rice+2016}, COHRS \citep[\twCO~J=3--2;][]{Colombo+2019}, CHIMPS \citep[\thCO\ and \CeiO~J=3--2;][]{Rigby+2019}, SEDIGISM \citep[\thCO~J=2--1;][]{Duarte-Cabral+2021}, and FUGIN \citep[\twCO, \thCO, and \CeiO~J=1--0;][]{Fujita+2023}.
Cloud radii are defined as the half width at half maximum (HWHM). 
In the middle and right panels, the MWISP point density is shown on a square-root scale. The other samples are represented by blue stars (GRS), yellow squares (CfA-Chile), brown crosses (COHRS), red plus signs (CHIMPS), green triangles (SEDIGISM), and magenta diamonds (FUGIN).
For clarity, only two hundred randomly selected clouds from each sample are displayed. 
Linear fits are performed on all well-resolved samples for the radius-mass and radius-velocity dispersion relations.
The parameters are in log form, therefore, the linear fit is equivalent to a power-law fit in the original variables.
The best-fit results are indicated by solid lines, and their slopes are noted in the upper-left corners of the respective panels.
}
\label{f:corrcomp}
\end{figure*}

The correlations between individual parameters are further examined.
The relationship between velocity dispersion ($\sigma_{\rm v}$) and cloud size ($L$), first formulated by \cite{Larson1981} as $\sigma_{\rm v} \propto L^\beta$ with $\beta$$\sim$0.38, is a cornerstone of MC studies. Our data allow us to make a direct comparison with this scaling law.
Figure~\ref{f:corrcomp} shows the radius distribution and the radius-velocity dispersion relation of our MC sample and those from other literature catalogs.
Well-resolved MC samples from other catalogs include: GRS MCs with physical properties estimated and masses greater than their uncertainties \citep{Roman-Duval+2010}; the CfA-Chile sample with parameters greater than their errors and sizes and velocity dispersions greater than their corresponding resolutions \citep{Rice+2016}; the COHRS fiducial sample with velocity dispersions greater than their errors and the velocity resolution \citep{Colombo+2019}; the CHIMPS highest reliability sample with parameters greater than their errors and velocity dispersions greater than the velocity resolution \citep{Rigby+2019}; the SDIGISM science sample with velocity dispersions greater than the velocity resolution \citep{Duarte-Cabral+2021}; and the FUGIN sample with distances greater than the 1~kpc error, positive mass values, and sizes and velocity dispersions greater than their corresponding resolutions \citep{Fujita+2023}.
The fitted $\sigma_{\rm v}$-$L$ relation here is relatively flat. 
We note that if the G50 positive-velocity MCs whose masses exceed the 1$\sigma$ scatter of the $\sigma_v$-M relation (see $kinematic distance$ estimation in Section~\ref{sec:para}) are included in the fit, the $\sigma_{\rm v}$-$L$ relation becomes even flatter.
This contributes to the ongoing discussion about the universality of Larson's relation \citep{Solomon+1987, CaselliMyers1995, Vazquez-Semadeni+1997, Heyer+2009, Simon+2001, Shetty+2012}. 
Studies have shown that the exponent $\beta$ can be shallower for cloud samples in the inner Galactic region \citep{Simon+2001, Shetty+2012}. 
It is noteworthy that \cite{Heyer+2001} found results similar to ours regarding the $\sigma_{\rm v}$-$L$ relation: small clouds show a flat relation, while large clouds ($\gtrsim9\,pc$) show a conventional steep relation \citep[see Figure~6 in][]{Heyer+2001}. 
These results suggest that some clouds may deviate from the global scaling, possibly because they are not fully virialized or are dominated by different physics. 
Our identification of a supra-Larson velocity dispersion MC sample (with $\sigma_{\rm v}$ greater than 1.5 times the fitted Larson relation) specifically highlights populations that deviate from the mean trend (see Section~\ref{sec:subsampprop}). These MCs are mostly nearby and small, with properties suggesting they may be young clouds that have inherited perturbations from the diffuse ISM. 

Another important empirical characterization of MC structure is the mass-size relationship, where $M \propto L^\beta$, with $\beta$$\sim$1.9 according to \cite{Larson1981}. This relationship was demonstrated by \cite{Zhou+2025} for a more complete cloud sample. Here, the cloud sample is more strictly selected and relatively larger. 
Figure~\ref{f:corrcomp} shows the mass-size relation of our MC sample and those from other literature catalogs.
For all catalogs, the mass and size are highly correlated and exhibit a good power-law relation. 
The FUGIN catalog contains a substantial number of MCs, nevertheless, the corresponding mass-size relation appears to be skewed by some extreme values. This issue requires further examination to exclude data points with unphysical parameter estimates.
The exponent of the mass-size relation derived here is consistent with those derived from other literature catalogs, except for the GRS, CHIMPS, and FUGIN catalogs.
We note that including the G50 positive-velocity MCs whose masses exceed the 1$\sigma$ scatter of the $\sigma_v$-M relation (see $kinematic distance$ estimation in Section~\ref{sec:para}) does not alter our fitting result.
The GRS traces relatively large \thCO\ clouds, and the CHIMPS catalog detects \thCO\ and \CeiO~(J=3--2) emission that primarily traces relatively dense gas, both of which have flatter mass-size relations.
Compared to the MC sample in \cite{Zhou+2025}, the MC sample here is well-resolved, containes fewer small MCs, and has a flatter mass-size relation.
These indicate that dense gas and large clouds have relatively lower space-filling factors. 

We also examine other correlations, including those between individual morphological and observational parameters, as well as those between individual morphological and physical parameters.
We perform linear fits for parameter pairs with a correlation coefficient greater than 0.6. Since the parameters are in log form, the fit is equivalent to a power-law fit, except for the morphology parameters ($R_2$, $\theta$, and $e$), distance ($D$), Galactocentric distance ($R_{\rm gal}$), and height from the Galactic plane ($z_{\rm gal}$). The fitting is weighted by the signal-to-noise ratio of the logarithm of the \twCO\ average integrated intensity, log~$\overline{W}$(\twCO). Tables of the fitted parameters of the correlations are available online in the Science Data Bank at doi:10.57760/sciencedb.32354 \citep{Zhou+2026data}, as are the corresponding correlation figures. 

\subsubsection{Dimensional Analysis and Eigenparameters}\label{sec:correigen}
\begin{figure*}[ptbh!]
\centerline{{\hfil\hfil
\psfig{figure=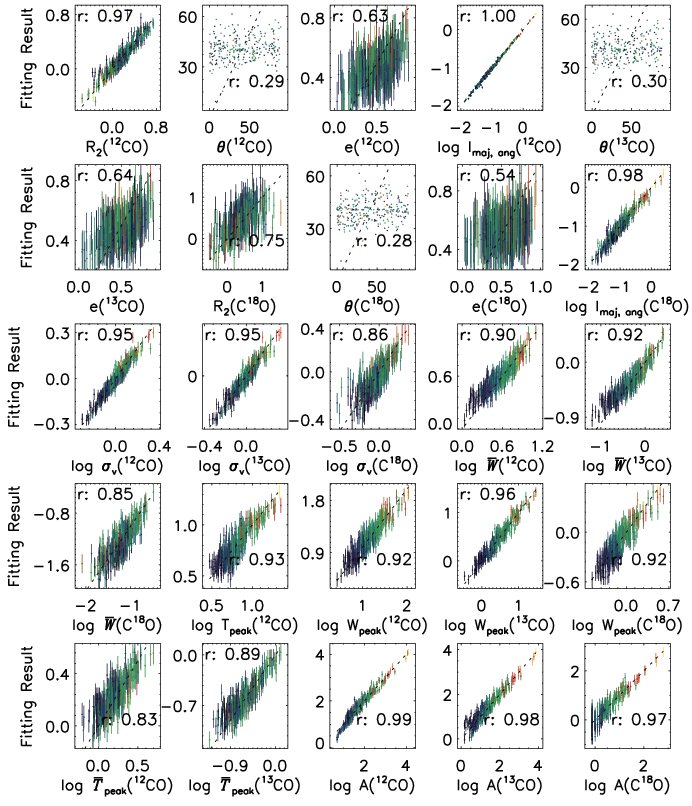,width=6in,angle=0, clip=}
\hfil\hfil}}
\caption{Fitting results of various observational parameters as a linear function of six selected observational eigenparameters.
The observational eigenparameters are $log\,V_{\rm span}$, $R_2$(\thCO), $log\,l_{\rm maj,\,ang}$(\thCO), $log\,T_{\rm peak}$(\thCO), $log\,T_{\rm peak}$(\CeiO), and $log\,\overline{T}_{\rm peak}$(\CeiO).
These parameters are in log form, so the fit is equivalent to a power-law fit to the original parameters, except for the morphological parameters ($R_2$ and $\theta$).
The fitting is weighted by the significance of $\overline{W}$(\twCO), indicated by colors from blue to red. The magnitude of $\overline{W}$(\twCO) is shown in its plot in the third row.
All fitted parameters are available in the Science Data Bank at doi:10.57760/sciencedb.32354 \citep{Zhou+2026data}. 
}
\label{f:corrfitpar}
\end{figure*}
\begin{figure*}[ptbh!]
\centerline{{\hfil\hfil
\psfig{figure=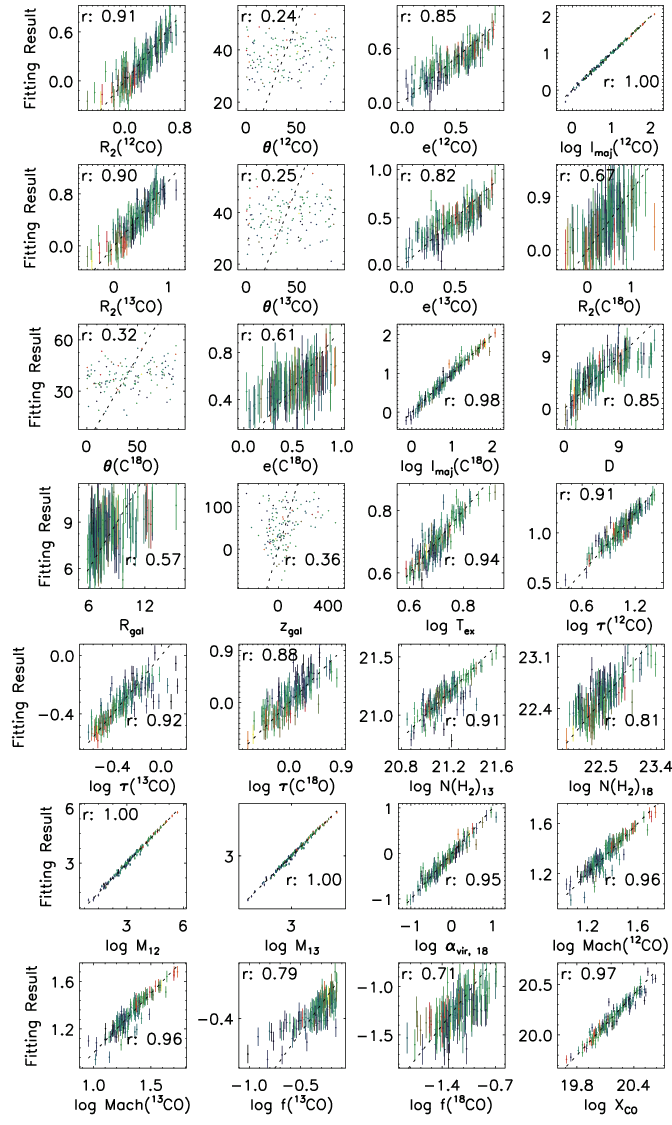,height=8.5in,angle=0, clip=}
\hfil\hfil}}
\caption{Same as Figure~\ref{f:corrfitpar}, but for derived physical parameters.
The physical eigenparameters are $log\,l_{\rm maj}$(\thCO), $log\,$\ncol$_{12}$, $log\,M_{18}$, $log\,\alpha_{\rm vir\,12}$, $log\,\alpha_{\rm vir\,13}$, and $log\,Mach$(\CeiO).
The physical parameters are in log form, except for the morphological parameters ($R_2$ and $\theta$), distance $D$, Galactocentric distance $R_{\rm gal}$, and height from the Galactic plane $z_{\rm gal}$.
All fitted parameters are available in the Science Data Bank at doi:10.57760/sciencedb.32354 \citep{Zhou+2026data}. 
}
\label{f:corrfitphy}
\end{figure*}

We perform a more detailed correlation analysis guided by dimensional analysis.
The intensity, velocity (derived from frequency via the Doppler shift effect), and spatial scale are the elementary quantities observed in the three CO lines. Among the interrelated observational parameters, a minimal set of independent eigenparameters consisting of these elementary quantities can be selected to estimate the magnitude of other parameters. The selection of this set is guided by the Buckingham $\Pi$ theorem of dimensional analysis \citep{Buckingham1914}. The theorem states that any physically meaningful relationship between $n$ dimensional variables can be expressed by a set of $n-k$ independent dimensionless functions ($\Pi$ groups), where $k$ is the number of fundamental dimensions involved.
In our case, there are nine elementary quantities, however, an elementary quantity of one CO line can be related to that of another CO isotopic line. Therefore, the number of fundamental dimensions is needed to be determined at first. 

There are at least three fundamental dimensions of observational parameters: intensity (in K), velocity (in \km\ps), and spatial scale (in degrees). A set of eigenparameters must span these dimensions to reconstruct of other parameters through power-law combinations. 
The chosen set of eigenparameters is not unique. Among different combinations, we select the one with low self-correlation (variance inflation facor $<$ 10), few eigenparameters, and one that can still fits the other parameters well (reject probabiliteis $<$ 0.05 significance level). 
The optimal number of eigenparameters that achieves the highest overall fit (highest mean adjusted coefficient of determination) is determined to be six.
The coefficient of determination ($R^2$) measures the proportion of variance explained by the fit and is calculated as $R^2=\Sigma(y_{fit,\,i}-\overline{y})^2/\Sigma(y_i-\overline{y})^2$, where $y_i$ is the parameter value, $y_{fit,\,i}$ is the fitted value of the parameter, and $\overline{y}$ is the mean of all parameter values.
While the adjusted coefficient of determination ($R_{\rm adjust}^2$) suppresses the addition of irrelevant eigenparameters and is defined as $R_{\rm adjust}^2=1-(1-R^2)(n-1)/(n-k-1)$, where n is the total number of parameters and k is the number of eigenparameters.
The selected set of observational eigenparameters is $log\,V_{\rm span}, R_2$(\thCO), $log\,l_{\rm maj,\,ang}$(\thCO), $log\,T_{\rm peak}$(\thCO), $log\,T_{\rm peak}$(\CeiO), and $\overline{T}_{\rm peak}$(\CeiO). This set effectively represents the scales (velocity span and angular sizes), the shape/density concentration (RJ-plot parameter $R_2$) and peak intensity measurements, and provides a good fit for the other parameters.

The linear fitting results of various observational parameters using these eigenparameters are shown in Figure~\ref{f:corrfitpar}. The linear fit of the logarithm of the parameters is equivalent to a power-law fit of the parameters themselves. All correlation coefficients $r$ between the observational parameters and their fitting results are greater than 0.7, except for the oblique angle $\theta$ and ellipcity $e$. The fit for the ellipcity is not as good as that of the other parameters, indicating a weaker correlation with the them.
The oblique angle $theta$, which indicates the influence of the Galactic disk gravity, is only strongly correlated with itself across the three CO isotopic lines (see Figure~\ref{f:matrix}). This self-correlation across tracers indicates that the influence of the Galactic gravity is a persistent property, independent of the density regime being traced and does not significantly affect other MC properties. 
Physical parameters are derived from the observational parameters and therefore have the same number of fundamental dimensions.
The selected eigenparameter set consists of $log\,l_{\rm maj}$(\thCO), $log\,N({\rm H)}_2)_{12}$, $log\,M_{18}$, $log\,\alpha_{\rm vir,\,12}$, $log\,\alpha_{\rm vir,\,13}$, and $log\,Mach$(\CeiO). 
Figure~\ref{f:corrfitphy} shows the linear fitting results for various derived physical parameters using these eigenparameters. Most of the derived physical parameters can be well-fitted, with correlation coefficients greater than 0.7. 
However, the oblique angle $\theta$, Galactocentric distance $R_{\rm gal}$, and Galactic height $z_{\rm gal}$ are poorly fitted, suggesting that the Galactic environment only has a limited influence on cloud properties in general.
The fit for the ellipticity and RJ-plot parameter of \CeiO\ $R_2$(\CeiO) is also poor, indicating that the collapse process of high-density gas is not strongly correlated with the overall properties of the cloud.

The importance of each eigenparameter is examined based on its mean normalized coefficient. 
This analysis reveals that the two most significant observational eigenparameters are the spatial scale $log\,l_{\rm maj,\,ang}$(\thCO) and the velocity scale $log\,V_{\rm span}$. The remaining eigenparameters are comparatively less important, in the order of $log\,T_{\rm peak}$(\thCO), $\overline{T}_{\rm peak}$(\CeiO), $R_2$(\thCO), and $log\,T_{\rm peak}$(\CeiO). 
Among the derived physical eigenparameters, the most important is the size $log\,l_{\rm maj}$(\thCO), followed by $log\,M_{18}$, $log\,\alpha_{\rm vir,\,12}$, $log\,\alpha_{\rm vir,\,13}$, $log\,Mach$(\CeiO), and $log\,N({\rm H)}_2)_{12}$.

\section{Summary} \label{sec:sum}
This study presents a comprehensive statistical analysis of MC properties using unbiased multi-tracer CO data (\twCO, \thCO, and \CeiO~J=1--0) from the MWISP survey, focusing on the inner (G50) and outer (G120) Galactic regions. From a parent catalog of 24,724 MCs, a final sample of 3,161 well-resolved clouds (including 324 with \CeiO\ emission) is established for detailed analysis of their observational, morphological, and physical parameters. The key findings are as follows:

1.\ Statistical Distributions and Galactic Influence\\
Most physical parameters follow log-normal distributions, while morphological parameters (ellipticity and RJ-plots parameters) follow normal distributions. MCs are typically oblate with a mean $e\sim0.46$, and their major axes tend to align parallel to the Galactic disk, indicating shaping by the disk's gravitational potential. However, this alignment weakens when local factors dominate, such as dense cores dominated by self-gravity.
The multi-peaked distributions of cloud distance and Galactocentric radius suggest an association with spiral arms.

2.\ Distance Selection Effects\\
Analysis of a nearby subsample ($D \le 4$~kpc) confirms that observational limits bias parameter distributions in the overall sample. At larger distances, the beam filling factor is lower, and only larger, brighter MCs are resolved, which skew the MC sample. A comparison study clarifies which parameter trends are intrinsic versus distance-induced.


3.\ Characteristics in Special Environments\\
Analysis of other subsamples reveals distinct properties. MCs possibly in the G120 spiral-shock region exhibit more concentrated/elongated \thCO\ gas and variable excitation temperatures.
MCs in the G50 interarm spurs are more elongated, have lower \thCO\ optical depth, and exhibit greater scatter in size and mass.
C18O-bright MCs are large, massive clouds with higher excitation temperatures and dense gas fractions.
Supra-Larson velocity dispersion MCs ($\sigma_{\rm v} > 1.5\times$Larson) are predominantly small, nearby clouds, which are possibly young and retain turbulence from the diffuse ISM.


4.\ Internal Structure from Multi-Tracer Analysis\\
Comparing parameters corresponding to the three CO lines reveals a schematic MC structure: \twCO\ traces diffuse, turbulent outer layers (turbulent timescale $<$ free-fall timescale), which do not contribute directly to star formation; \thCO\ traces a transitional regime; and C18O traces dense, gravitationally dominated cores (virial parameter $\lesssim$ 1) where collapse can proceed.

5.\ Scaling Relations and Dimensional Analysis\\
Correlation analysis confirms known relations (e.g., mass-size) but reveals a flatter velocity dispersion-size relation than classic Larson's law, suggesting the presence of a population of non-virialized clouds or different governing physics. A full dimensional analysis identifies minimal sets of eigenparameters, from which most other parameters can be estimated via power-law combinations. 
The observational eigenparameters, ranked by importance, are $log\,l_{\rm maj,\,ang}$(\thCO), $log\,V_{\rm span}$, $log\,T_{\rm peak}$(\thCO), $\overline{T}_{\rm peak}$(\CeiO), $R_2$(\thCO), and $log\,T_{\rm peak}$(\CeiO). The physical eigenparameters, ranked by importance, are $log\,l_{\rm maj}$(\thCO), $log\,M_{18}$, $log\,\alpha_{\rm vir,\,12}$, $log\,\alpha_{\rm vir,\,13}$, $log\,Mach$(\CeiO), and $log\,N({\rm H)}_2)_{12}$.
In addition to the size, velocity dispersion, mass, and virial parameters that are related to Larson's relations, the uniformity and H$_2$ column density are also essential properties. 
In conclusion, these findings advance our understanding of the statistical properties, environmental dependencies, and internal structure of Galactic MCs. 
Future high-resolution and high-sensitivity surveys or the combined analysis of multi-line survey data can better determine the physical state of molecular gas and further clarify the evolution and scaling relations of MCs.

\begin{acknowledgments}
This research made use of the data from the Milky Way Imaging Scroll Painting (MWISP) project, which is a multi-line survey in \twCO/\thCO/\CeiO\ along the northern Galactic plane with PMO-13.7m telescope. We are grateful to all the members of the MWISP working group, particularly the staff members at PMO-13.7m telescope, for their long-term support. MWISP was sponsored by National Key R\&D Program of China with grants 2023YFA1608000 \& 2017YFA0402701 and by CAS Key Research Program of Frontier Sciences with grant QYZDJ-SSW-SLH047.
X.Z. acknowledges support from the National SKA Program of China (2025SKA0140100).
\end{acknowledgments}
\paragraph{Data Availability}
All distribution and correlation figures of MC parameters, along with the complete set of corresponding fitted parameters (including those for the eigenparameter-based relations), are available in the Science Data Bank at doi:10.57760/sciencedb.32354 \citep{Zhou+2026data}. 
 
\bibliographystyle{aasjournal}


\end{document}